\documentclass[notitlepage, superscriptaddress
,reprint 
]{revtex4-1} 

\usepackage[utf8]{inputenc}
\usepackage{graphicx}
\usepackage{xcolor}
\usepackage{float}
\usepackage{amsmath}
\usepackage{amssymb}
\usepackage{braket}
\usepackage{bm}
\usepackage{hyperref}
\usepackage{bbm}
\usepackage{bbold}
\usepackage[normalem]{ulem}
\newcommand{\sinc}{\mathrm{sinc}}

\newcommand{\Hc}{\mathcal{H}}
\newcommand{\Dc}{\mathcal{D}}
\newcommand{\Jc}{\mathcal{J}}

\newcommand{\ca}{$^{40}$Ca$^+$}

\begin{document}

\title{Exactly solvable model of light-scattering errors in quantum simulations with metastable trapped-ion qubits}
\author{Phillip C. Lotshaw}
\email[]{Lotshawpc@ornl.gov}
\affiliation{Quantum Information Science Section, Oak Ridge National Laboratory, Oak Ridge, TN 37381, USA}

\thanks{This manuscript has been authored by UT-Battelle, LLC, under Contract No. DE-AC0500OR22725 with the U.S. Department of Energy. The United States Government retains and the publisher, by accepting the article for publication, acknowledges that the United States Government retains a non-exclusive, paid-up, irrevocable, world-wide license to publish or reproduce the published form of this manuscript, or allow others to do so, for the United States Government purposes. The Department of Energy will provide public access to these results of federally sponsored research in accordance with the DOE Public Access Plan.}

\author{Brian C. Sawyer}
\affiliation{Georgia Tech Research Institute, Atlanta, GA 30332, USA}

\author{Creston D. Herold}
\affiliation{Georgia Tech Research Institute, Atlanta, GA 30332, USA}

\author{Gilles Buchs}
\affiliation{Quantum Information Science Section, Oak Ridge National Laboratory, Oak Ridge, TN 37381, USA}

\date{\today} 

\begin{abstract}
    We analytically solve a model for light scattering in Ising dynamics of metastable atomic qubits, generalizing the approach of Foss-Feig {\it et al.}~[Phys.~Rev.~A {\bf 87}, 042101 (2013)] to include leakage outside the qubit manifold. We analyze the influence of these fundamental errors in simulations of proposed experiments with metastable levels of $^{40}$Ca$^+$ ions in a Penning trap. We find that ``effective magnetic fields" generated by leaked qubits have significant impacts on spin-spin correlation functions for Greenberger-Horne-Zeilinger state preparation or for quantum simulations with strong coupling, while spin squeezing uses a much weaker coupling and is largely insensitive to the simulated leakage errors, even with a few hundred ions. Our theory and results are expected to be useful in modeling a variety of metastable qubit experiments in the future.
\end{abstract}
\maketitle

\emph{Introduction}.~Metastable atomic qubits offer multiple advantages over more traditional ground or optical qubit encodings. Their use expands the range of available wavelengths for qubit control, notably increasing the wavelength of optical controls relative to those of ground state qubits \cite{Allcock_APL:2021}, and a practical advantage is the availability of high-power near-infrared lasers for metastable qubit operations. However, metastable qubits also have distinct fundamental errors due to photon-scattering pathways that are not present in ground state encodings. To better understand these errors we solved a model of light scattering dynamics in metastable-qubit quantum simulations.

\begin{figure}[ht]
    \includegraphics[height=8cm,width=8.5cm,keepaspectratio]{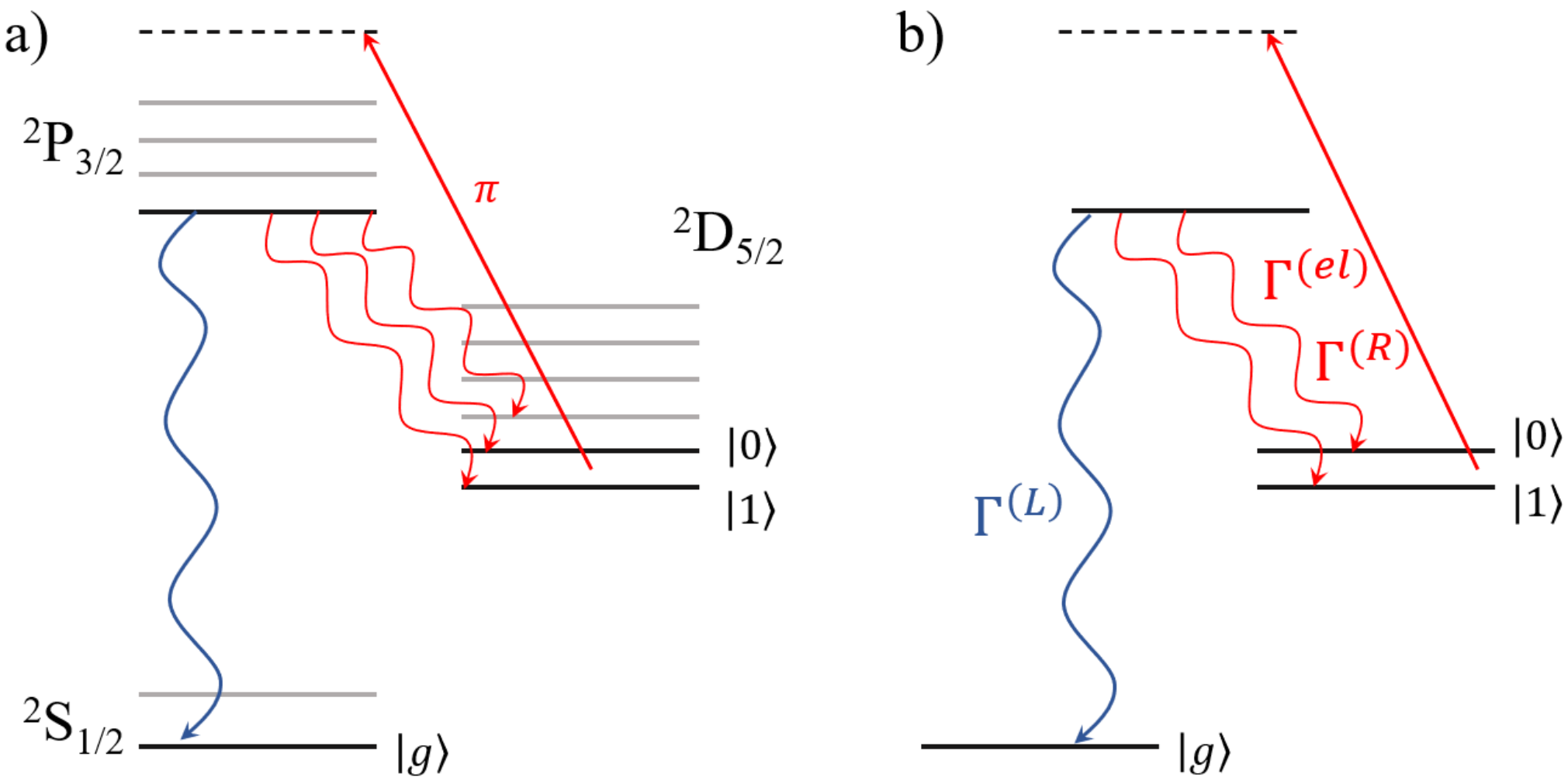}
    \caption{(a) Partial electronic energy level diagram of \ca~including metastable qubit levels $\ket{0}$ and $\ket{1}$ in the $^2\text{D}_{5/2}$ manifold. A $\pi$-polarized far-detuned laser beam (straight arrow) generates AC Stark shifts of the qubit levels through their interaction with the indicated $^2\text{P}_{3/2}$ level (black line). Off-resonant light scattering generates errors through allowed decay pathways (wavy lines). (b) Simplified level diagram used for analytic calculations with generic metastable qubits. The $\Gamma^{(R)}$, $\Gamma^{(el)}$, and $\Gamma^{(L)}$ are rates of Raman scattering in the qubit manifold, elastic scattering, and leakage to the ground state $\ket{g}$, respectively.}
    \label{level diagram}
\end{figure}

Our work builds on a previous analysis of Foss-Feig \emph{et al.}~\cite{Foss-Feig_PRA:2013}, who analytically solved a model for non-equilibrium dynamics of an Ising interaction in which spontaneous photon scattering leads to fluctuating spins which mimic a noisy magnetic field. Their model gave a good account of decoherence in quantum simulations with ${}^9\mathrm{Be}^+$ ions in the NIST Penning trap system, which has been used to demonstrate spin squeezing of hundreds of ions \cite{Bohnet_Science:2016}, out-of-time-order correlations in spin magnetization \cite{Garttner_Nature_Phys:2017}, and force sensing \cite{gilmore2021sensing}. While scattering in ${}^9\mathrm{Be}^+$ ground-level qubits remained confined to the qubit manifold, here we consider metastable qubits such as the $^2\text{D}_{5/2}$ levels of ${}^{40}\mathrm{Ca}^+$ \cite{mcmahon_individual_2024} depicted as $\ket{0}$ and $\ket{1}$ in Fig.~\ref{level diagram}(a). Light scattering from these qubit levels predominantly induces leakage to the ground state $\ket{g}$, which can be detected without disturbing the metastable qubit. Unlike scattering errors within the qubit manifold, detecting these leakage errors has benefits for quantum error correction schemes utilizing ``erasure conversion" \cite{wu_erasure_2022, kang_erasure_2023}. Quantum simulation accuracy can also be improved by detecting and discarding results with leakage, at the cost of additional sampling overhead, as  we quantify below.  

Here we analytically solve a model for decoherent Ising dynamics with light scattering and leakage, with generic metastable qubits as illustrated in Fig.~\ref{level diagram}(b). We examine the influence of light scattering errors in proposed experiments with \ca (Fig.~\ref{level diagram}(a)), where we include an additional analytic treatment of a spin-echo sequence.  We find differences in leakage error sensitivity for strongly-coupled quantum spin dynamics and Greenberger-Horne-Zeilinger (GHZ) state preparation, in contrast to spin squeezing and a simple single-layer benchmarking implementation of the quantum approximate optimization algorithm (QAOA). Our results are expected to be useful in analyzing future experiments with metastable trapped-ion qubits.

\emph{Light shift interaction}.~Applying two off-resonant laser beams to $N$ trapped ions generates an instantaneous optical dipole force Hamiltonian 
\begin{equation}
    \label{Ht} H(t) = \sum_{i=1}^N F_0 \cos(\mu t) \hat z_i \sigma^z_i
\end{equation}
where $F_0$ is the force amplitude and $\mu$ is the laser beatnote angular frequency \cite{Bohnet_Science:2016,Sawyer_PRA:2021}. Here $\sigma^z_i$ represents the Stark shift to the electronic energy levels of ion $i$, coupled to the vibrational states through the position operator $\hat z_i$. This ``spin-dependent force" causes vibrational modes to traverse loops in phase space with periods $T_m = 2\pi/\delta_m$ where $\delta_m=\mu-\omega_m$ is the detuning from resonance with respect to mode $m$, with angular frequency $\omega_m$ \cite{Sawyer_PRA:2021}.  When a mode returns to its origin it becomes disentangled from the electronic levels, leaving a residual geometric phase which is exploited for quantum information processing. These phases yield an effective Ising evolution $\exp(-i \Hc t)$ after integrating (\ref{Ht}) in the Lamb-Dicke approximation and using a spin-echo sequence to cancel undesired ``longitudinal field" terms $\sim \sigma^z_i$ \cite{monroe2021programmable,Sawyer_PRA:2021}, where
\begin{equation} \label{HIsing} \Hc = \frac{1}{N}\sum_{i<j} J_{ij} \sigma^z_i \sigma^z_j \end{equation}
with
\begin{equation} \label{Jij} J_{ij} = N\Omega_i \Omega_j \sum_{m}\frac{\eta_{i,m} \eta_{j,m}\omega_m}{\mu^2-\omega_m^2} \end{equation}
where $\eta_{i,m}$ is a Lamb-Dicke parameter.

\emph{Ising dynamics with light scattering and leakage}.~Uys \emph{et al.~}\cite{uys2010decoherence} derived a Lindbladian master equation that accurately described single-ion light-scattering errors in $^9$Be$^+$.  Foss-Feig \emph{et al.}~\cite{Foss-Feig_PRA:2013,fossfeigdissertation} applied the same error model to the effective Ising evolution (\ref{HIsing}) and, remarkably, obtained an analytic formula that successfully described decoherence in spin-squeezing experiments with hundreds of ions \cite{Bohnet_Science:2016}. We generalize this approach to include leakage [Fig.~\ref{level diagram}(b)].

We consider an experimentally-relevant \cite{leibfried2005creation,pezze2018quantum,rajakumar2022generating} initial state $\ket{\psi_0} = \ket{+}^{\otimes N}$ evolving under the Lindbladian master equation \cite{uys2010decoherence,Foss-Feig_PRA:2013}
\begin{equation}
   \label{master equation} \dot \rho = -i(\Hc_\text{eff}\rho - \rho \Hc_\text{eff}^\dag) +  \Dc(\rho) 
\end{equation}
where the effective non-Hermitian Hamiltonian $\Hc_\text{eff}$ and dissipator $\Dc(\rho)$ are
\begin{equation} \Hc_\text{eff} = \Hc -i\sum_{\alpha,j} {\Jc^{(\alpha)}_j}^\dag\Jc^{(\alpha)}_j,\ \ \  \Dc(\rho) =2\sum_{\alpha,j} \Jc^{(\alpha)}_j \rho {\Jc^{(\alpha)}_j}^\dag \end{equation} 
Jump operators $\Jc^{(\alpha)}_j$ are included for each type of scattering event indicated in Fig.~\ref{level diagram}(b), including elastic scattering with $\Jc^{(el)}_j = \sqrt{\Gamma^{(el)}/8}\sigma_j^z$ as well as inelastic scattering with $\Jc^{(a\to  b)}_j = \sqrt{\Gamma^{(a \to b)}/2}\ket{b_j}\bra{a_j}$ for $a_j \in \{0,1\}, b_j\in \{0,1,g\}$ with $b_j \neq a_j$. The latter transitions include Raman scattering $\ket{0}\leftrightarrow\ket{1}$ within the qubit manifold, with the single-ion Raman decoherence rate $\Gamma^{(R)} = (\Gamma^{(0\to1)} + \Gamma^{(1\to0)})/2$, and leakage scattering to $\ket{g}$, with the single-ion leakage decoherence rate $\Gamma^{(L)} = (\Gamma^{(0\to g)} + \Gamma^{(1\to g)})/2$. The total single-ion decoherence rate is $\Gamma =\Gamma^{(L)} + \Gamma^{(R)} + \Gamma^{(el)}/2$.

We compute exact dynamics of (\ref{master equation}) using the ``quantum trajectories" approach \cite{Foss-Feig_PRA:2013,fossfeigdissertation,carmichael2009open}.  The basic idea is to organize the evolution (\ref{master equation}) into pure-state ``trajectories" which are easier to analyze and are summed to produce the exact time-dependent state. The first term in (\ref{master equation}) generates ideal evolution, with probability less than one.  The second term $\Dc(\rho)$ produces quantum jumps under light scattering. 

A key insight from Ref.~\cite{Foss-Feig_PRA:2013} is that if a jump operator such as $\Jc^{(0\to 1)}_i \sim \ket{1_i}\bra{0_i}$ is applied to an otherwise ideal state $\ket{\psi}$ at time $t_i$, this will generate a transition $\ket{0_i} \to \ket{1_i}$ and remove any part of the wavefunction that was in $\ket{1_i}$ before the jump occured.  Hence, it will be as if the ion began in the state $\ket{0_i}$, which is an eigenstate of $\sigma^z_i$ in the Ising interaction (\ref{HIsing}). Replacing $\sigma^z_i$ with the corresponding eigenvalue leads to an ``effective magnetic field" $N^{-1}\sum_{j \neq i} J_{ij}\sigma_i^z$ for time $t<t_i$ (for the state $\ket{0_i}$) and $-N^{-1}\sum_{j \neq i} J_{ij}\sigma_i^z$ for $t \geq t_i$ (after scattering to $\ket{1_i}$); note that we are using a convention in which $\ket{1_i}$ has lower energy than $\ket{0_i}$, see~Fig.~\ref{level diagram}. In the case of multiple scattering events, the effective field strength depends on the difference in times $t_0-t_1$ that the ion spends in $\ket{0_i}$ and $\ket{1_i}$. This idea was used previously to derive exact correlation functions including all possible Raman scattering processes within the qubit manifold \cite{Foss-Feig_PRA:2013}.  The same idea remains essential in our treatment of leakage, which generates an effective field that ``turns off" when an ion scatters to $\ket{g_i}$.

From (\ref{master equation}) we derived a complete set of time-dependent operator expectation values $\langle \bm \sigma \rangle$ where $\bm \sigma$ is any product of raising $\sigma^+_i = \ket{0_i}\bra{1_i}$ or lowering $\sigma^-_i=\ket{1_i}\bra{0_i}$ operators, or single-level projectors $\sigma^z_i = \ket{z_i} \bra{z_i}$, acting on any arbitrary set of qubits. There is no coherence between the ground and qubit levels in our model, so operators describing these coherences (such as $\ket{g_i}\bra{0_i}$) have zero expectation.  The $\bm \sigma$ form a basis for the relevant Liouville operator space, hence any operator $O$ can be expressed as $O = \sum_{\bm \sigma} O_{\bm \sigma} \bm \sigma$ and its expectation can be computed as $\langle O \rangle = \sum_{\bm \sigma} O_{\bm \sigma} \langle \bm \sigma \rangle$. We can also compute $\rho(t)$ analytically. This generalizes the one- and two-spin analysis of Foss-Feig {\it et al.}~\cite{Foss-Feig_PRA:2013} by extending to arbitrary many-body operators and by including leakage as a new error source.  We verified these analytic results against direct numerical integrations of (\ref{master equation}) using Simpson's method \cite{numericalrecipes}, with arbitrarily chosen nonuniform couplings and scattering rates, with up to four ions evolving for thousands of small timesteps.  We verified that our theory was correct to numerical precision for several physical quantities, including all matrix elements of the time evolving state $\rho(t)$. Detailed mathematical expressions and derivations are in Supplementary Information (SI) Sec.~A and B.

As an example, we consider spin-spin correlation functions involving $\sigma^x_i  = \sigma^+_i + \sigma^-_i$  and $\sigma^y_j = -i(\sigma^+_i-\sigma^-_j)$. To compute an $m$-body correlation function such as $\langle \prod_{j \in M}\sigma^x_j\rangle$, over a set of ions $M = \{j_1, j_2, \ldots, j_m\}$, we expand each $\sigma^x_j$ in terms of $\sigma^+_j$ and $\sigma_j^-$ to obtain a sum of $2^m$ expectation values $\langle \prod_{j \in M} \sigma_j^{\nu_j} \rangle$ with $\nu_j \in \{+,-\}$. For example, $\langle \sigma^x_i\sigma^x_j\rangle=\langle \sigma^+_i\sigma^+_j\rangle + \langle \sigma^+_i\sigma^-_j\rangle + \langle \sigma_i^-\sigma^+_i\rangle + \langle \sigma^-_i\sigma^-_j\rangle$, with similar expressions for $\langle \sigma^x_i\sigma^y_j\rangle$ and $\langle \sigma^y_i\sigma^y_j\rangle$. In SI we find that each of the $\langle \prod_{j \in M} \sigma_j^{\nu_j} \rangle$ can be expressed as
\begin{align} \label{main result}
\left\langle \prod_{j \in M} \sigma_j^{\nu_j} \right\rangle = & \frac{e^{-m\Gamma t}}{2^m}  \prod_{i\notin M}  \bigg[ I\left(J_i^{(M,\bm \nu)} ,t\right) + R\left(J_i^{(M,\bm \nu)} ,t\right)  \nonumber\\
& + L\left(J_i^{(M,\bm \nu)} ,t\right) + B\left(J_i^{(M,\bm \nu)} ,t\right)\bigg], 
\end{align}
where the functions $I,R,L,B$ are defined below and $J_i^{(M,\bm \nu)} = \sum_{j \in M} \nu_j J_{ij}$, with the notation $\nu_jJ_{ij}=J_{ij}$ for $\nu_j=+$ and $\nu_jJ_{ij}=-J_{ij}$ for $\nu_j=-$.

The prefactor $\exp(-m \Gamma t)$ is the product of single-ion decoherences for the $m$ ions $j \in M$.  Subsequent terms describe the influence of other ions through the Ising interaction, which is accounted for in terms of specific types of trajectories.  

The first term 
\begin{align} I(J_{i}^{(M,\bm \nu)},t) = e^{- \lambda t} \cos(st). \end{align}
corresponds to ideal evolution, with inelastic relaxation rate $\lambda = (\Gamma^{(0 \to 1)} + \Gamma^{(0 \to g)} + \Gamma^{(1 \to 0)} + \Gamma^{(1 \to g)})/2$ and the effective damped oscillation frequency $s=s(J_i^{(M,\bm \nu)}) = 2J_i^{(M,\bm \nu)}/N + i\Delta$, which depends on the state-dependent scattering rate difference $\Delta =(\Gamma^{(0 \to 1)} + \Gamma^{(0 \to g)} - \Gamma^{(1 \to 0)} - \Gamma^{(1 \to g)})/2 $.

The second term 
\begin{align} R(J_{i}^{(M,\bm \nu)},t) = & e^{-\lambda t}\bigg[\cos(\zeta t) - \cos(st) + \Gamma^{(R)}t\, \mathrm{sinc}(\zeta t) \bigg]\end{align}
corresponds to trajectories under sequences of Raman transitions that do not leave the qubit manifold, with effective damped frequency $\zeta=\zeta\left(J_i^{(M,\bm \nu)}\right) = \sqrt{s^2(J_i^{(M,\bm \nu)} ) - \Gamma^{(0 \to 1)}\Gamma^{(1\to 0)}}$. Equivalent expressions for $I(J_i^{(M,\bm \nu)},t)$ and $ R(J_i^{(M,\bm \nu)},t)$ were derived previously in Eq.~(10) of Ref.~\cite{Foss-Feig_PRA:2013}.

The third term
\begin{align} L(J_{i}^{(M,\bm \nu)},t) = \frac{1}{2}\sum_{\alpha=0}^1\Gamma^{(\alpha \to g)} f(\Gamma^{(\alpha)},(-1)^\alpha2J_i^{(M,\bm \nu)} /N,t)\end{align}
describes leakage outside the qubit manifold, where $\Gamma^{(0)}= \Gamma^{(0\to g)} + \Gamma^{(1\to g)}$ and $\Gamma^{(1)} = \Gamma^{(1\to g)} + \Gamma^{(1\to 0)}$ are the total scattering rates from $\ket{0}$ and $\ket{1}$ respectively, and $f(\Gamma,\chi,t)=  e^{ i(\chi + i\Gamma)t/2} t\, \mathrm{sinc}[(\chi + i\Gamma)t/2]$. 

The final term
\begin{align} \label{B} B(J_{i}^{(M,\bm \nu)},t) = & \Gamma^{(L)}[a(\lambda,\zeta,t) - a(\lambda,s,t)]+ \Gamma^{(B)}b(\lambda,\zeta,t)\nonumber\\ 
& + is\Delta^{(L)}[b(\lambda,\zeta,t)-b(\lambda,s,t)] \end{align}
describes trajectories with both a sequence of Raman transitions and a final leakage transition, with $a(\Gamma,\chi,t) =  [f(\Gamma,\chi,t) + f(\Gamma,-\chi,t)]/2$,   
$b(\Gamma,\chi,t) = -i[f(\Gamma,\chi,t) - f(\Gamma,-\chi,t)]/2\chi$, $\Gamma^{(B)}=(\Gamma^{(0\to 1)}\Gamma^{(1\to g)} + \Gamma^{(1\to 0)}\Gamma^{(0\to g)})/2$ and $\Delta^{(L)} = (\Gamma^{(0\to g)}-\Gamma^{(1\to g)})/2$.

\emph{Metastable qubits in $^{40}$Ca$^+$}.~We base our model for photon scattering errors in \ca~metastable qubits on the compact Penning trap system operated at GTRI~\cite{mcmahon_individual_2024}. It exhibits a 0.9~T magnetic field, which splits the neighboring $D_{5/2}$ magnetic sublevels of \ca~ions by $\sim15$~GHz. Off-resonant laser beams with wavelengths near the 854~nm $D_{5/2} \rightarrow P_{3/2}$ transition are applied to induce a differential AC Stark shift between two metastable electronic states. As illustrated in Fig.~\ref{level diagram}(a), we define the qubit states as $\ket{1}\equiv \ket{D_{5/2}, m_J = -5/2}$ and $\ket{0}\equiv \ket{D_{5/2}, m_J = -3/2}$. A linearly-polarized infrared laser beam is applied such that only $\pi$ transitions are coupled, thereby inducing an AC Stark shift on only the $\ket{0}$ state of (in units of $s^{-1}$)
\begin{equation}
\Omega^{(0)} = \frac{6}{5} \frac{A_{P32D52}}{\hbar c k^3_{DP} \Delta_{P_{3/2}}} \left(\frac{P}{w_0^2}\right),
\end{equation}
where $\hbar$ is Planck's constant, $c$ is the speed of light, $A_{P32D52} \sim 8.48\times10^6~s^{-1}$ is the spontaneous decay rate from $P_{3/2}$ to $D_{5/2}$~\cite{gerritsma_precision_2008}, $k_{DP}$ is the magnitude of the transition k-vector, $\Delta_{P_{3/2}}\sim2\pi\times1$~THz is the laser beam detuning from resonance, and $P$ ($w_0$) is the total power (waist) of the shifting laser beam. Scattering rates are~\cite{wineland_quantum_information_2003}
\begin{equation} \Gamma^{(0 \to b)} = A_{P_{3/2} \to b} \left|\frac{\Omega^{(0)}}{\Delta_{P_{3/2}}}\right| \end{equation}
where $A_{P_{3/2} \to b}$ is the spontaneous decay rate from the $P_{3/2}$ sublevel in Fig.~\ref{level diagram}(a) to a final state $b$. We find total single-ion scattering rates of $< 11 $~s$^{-1}$ for simulations in later sections. When scattering occurs, leakage accounts for $94.5\%$ of the spontaneous decays, while elastic and Raman scattering account for the remaining $1.6\%$ and $3.9\%$, respectively~\cite{gerritsma_precision_2008}. Selection rules forbid scattering from $\ket{1}$ under our chosen polarization. 

With this setup we generate an effective Hamiltonian $\Hc_\text{arm} = N^{-1}\sum_{i<j} J_{ij} \ket{0_i 0_j}\bra{0_i0_j}$ under the light-shift interaction. Ising dynamics (\ref{HIsing}) are generated by applying $\Hc_\text{arm}$ for time $t_\text{arm}$ in each of two arms of a spin-echo sequence, yielding terms $\ket{0_i0_j}\bra{0_i0_j} + \ket{1_i1_j}\bra{1_i1_j} = (\sigma^z_i \sigma^z_j+1)/2$; we treat this spin-echo sequence analytically for the $^{40}$Ca$^+$ calculations that follow, see SI Sec.~C for details. Note that $t_\text{arm}$ is the relevant scattering timescale and is twice as long as the effective Ising evolution time $t$. We consider detuning $\delta \equiv \delta_0 \ll \delta_{m \neq 0}$ close to the center-of-mass (COM) $m=0$ mode to obtain nearly-equal couplings $J_{ij}/N$, and set $t_\text{arm} = 2\pi/\delta$ so that the COM mode returns to its initial state at the end of each spin-echo arm.  We compute the coupling elements $J_{ij}/N$ using zero-temperature simulations of equilibrium ion-crystal configurations as in~\cite{wang_phonon-mediated_2013}.

\emph{GHZ state preparation}.~As a first application of our model, we examine the preparation of maximally-entangled $N$-ion GHZ or ``Schr\"odinger cat" states \cite{greenberger1989going,mermin1990extreme}. An Ising Hamiltonian with uniform coupling $J/N$ produces a GHZ state at the cat time $t_\text{cat}=\pi N /4J$
\begin{equation} \label{GHZ} \ket{\psi_\text{GHZ}} = e^{-i JN^{-1} t_\text{cat} \sum_{i<j} \sigma^z_i \sigma^z_j} \ket{+}^{\otimes N} = \frac{\ket{a}^{\otimes N} + e^{i\theta} \ket{b}^{\otimes N}}{\sqrt{2}} \end{equation}
where $\ket{a} \neq \ket{b}$ are eigenstates of $\sigma^x_i$ for $N$ even or $\sigma^y_i$ for $N$ odd \cite{molmer1999multiparticle,leibfried2005creation}. 
In practice, the uniformity of the $J_{ij}$ is limited by practical detunings from the COM mode; a smaller detuning generates more uniform couplings at the cost of a slower operation.

We use the fidelity $F = \bra{\psi_\text{GHZ}} \rho(t_\text{cat}) \ket{\psi_\text{GHZ}}$ to quantify success in GHZ state preparation. Directly calculating $F$ is infeasible at large $N$ because of the exponential size of $\rho(t_\text{cat})$, so we estimate $F$ as a product of two parts which are easier to compute.

First, we assume all Ising couplings are equal to their average value $J_{ij}=J$, so $\rho$ can be computed in polynomial time using permutation invariance symmetry \cite{gegg2016efficient}.  This symmetrized state gives an estimate $F_\text{scatter}$ for how light scattering affects the fidelity. We calculated the symmetrized state $\rho(t_\text{cat})$, and hence $F_\text{scatter}$, using our theory; this included the effects of elastic, inelastic, and leakage scattering, and required only minimal computational resources compared to a direct numerical integration of (\ref{master equation}). Second, we use analytic approximations to treat the additional error $F_\text{unequal}$ due to coupling variations $\delta J_{ij} = J-J_{ij} \neq 0$ which drive $\rho$ away from the ideal GHZ state (\ref{GHZ}). At second order in the $\delta J_{ij}$ and for large $N$
\begin{equation} \label{Funequal} F_\text{unequal} \approx \exp\left[- t_\text{cat}^2 \sigma^2(J_{ij})/2\right],\end{equation}
where $\sigma^2(J_{ij})$ is the variance of the $J_{ij}$.  We evaluate $F_\text{unequal}$ to fourth order in results that follow; the derivation leading to $F_\text{unequal}$ is generic and independent of the specific light scattering model (\ref{master equation}), with details presented in SI Sec.~D. Our total fidelity estimate is $F \approx F_\text{unequal}F_\text{scatter}.$ We will also consider situations in which measurements are used to detect and discard runs with leakage. This improves the scattering fidelity to $F_\text{scatter} \approx 1$, with small residual error due to elastic and  Raman scattering.

\begin{figure}
    \centering
    \includegraphics[height=12cm, width=\columnwidth,keepaspectratio]{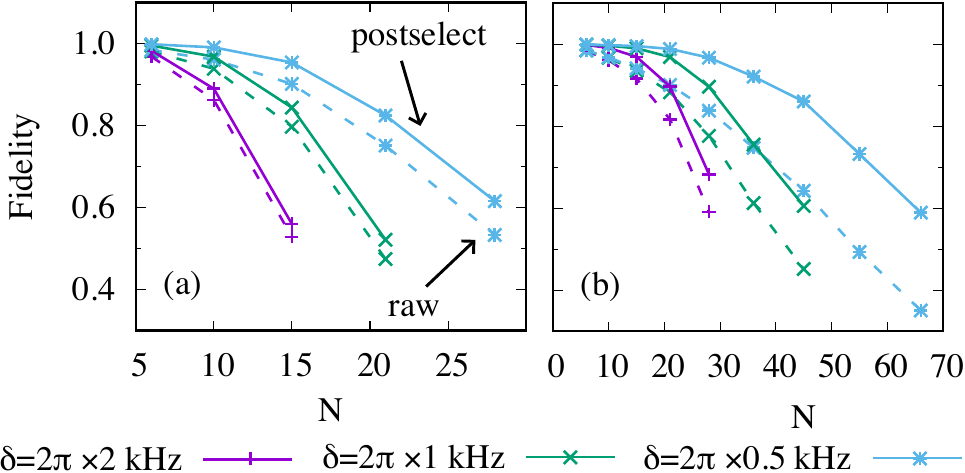}
    \caption{GHZ state preparation fidelity with (a) $B = 0.9$~T magnetic field and (b) $B=4.5$~T. Dashed lines are raw results while solid lines assume postselection to remove leakage.}
    \label{fig:GHZ}
\end{figure}

Figure \ref{fig:GHZ}(a) shows our estimated $F$ for GHZ states prepared in a 0.9 T magnetic field.  Each curve terminates when larger sizes would result in $F < 0.5$.  The dashed lines illustrate the raw $F$ while solid lines illustrate $F_\text{postselect}$ after postselection is used to remove leakage. Postselection yields a notable improvement, though $F_\text{postselect} \approx F_\text{unequal}$ is ultimately limited by the use of unequal couplings $J_{ij}$. 

Postselection requires an expected sampling overhead $1/P_\text{no\ leak}$ where $P_\text{no\ leak}$ is the probability of no leakage. We can compute this from the fidelities in Fig.~\ref{fig:GHZ} by considering that the raw $F$ is an average over trajectories with and without leakage, $F = P_\text{no\ leak} F_\text{no\ leak} + P_\text{leak}F_\text{leak} = P_\text{no\ leak}F_\text{no\ leak}$, where the last equality uses $F_\text{leak}=0$ since the GHZ state has zero overlap with the ground state subspace.  After postselection the fidelity improves to $F= F_\text{no\ leak}$. Hence, the expected sampling overhead $1/P_\text{no\ leak} = F_\text{postselect}/F$ and this is close to one for the results in Fig.~\ref{fig:GHZ}(a).

To understand prospects for GHZ preparation at larger sizes, in Fig.~\ref{fig:GHZ}(b) we increase the strength of the magnetic field to 4.5 T, which spreads the vibrational mode spectrum to decrease the variance $\sigma^2(J_{ij})$.  We estimate that GHZ states can be produced at approximately twice the sizes seen previously in Fig.~\ref{fig:GHZ}(a),  with greater gains from postselection. Another approach to reducing the spread in $J_{ij}$ could be to engineer a more uniform coupling using a multi-tone optical dipole force that simultaneously drives all axial motional modes \cite{shipira_theory_2020}. Alternately, dispersive measurements have been proposed as a non-deterministic means to generate large GHZ states with high fidelity \cite{alexander_generating_2020}.

\emph{Correlation functions}.~Quantum simulations often rely on ``local" or few-body observables, which may be much less sensitive to noise than the global state fidelity. Here we quantify the accuracy of $m$-body spin-spin correlation functions. GHZ states (\ref{GHZ}) constitute ideal standards for this analysis, since GHZ states polarized along $P \in \{\sigma^x,\sigma^y\}$ will satisfy $\langle P^{\otimes 2m}\rangle = 1$ for any even number of ions $2m$. To compare against this standard we analyze spin correlation functions assuming an equal coupling $J_{ij}=J$, so that any deviation $\langle P^{\otimes 2m}\rangle \neq 1$ is attributable directly to light-scattering errors. We expect similar physics in other cases, such as GHZ state preparation with unequal $J_{ij}$ and quantum simulations with strong couplings, with the additional complication that the noiseless reference $\langle P^{\otimes 2m}\rangle \neq 1$ will depend on details of the $J_{ij}$ and may therefore be much more variable.  

\begin{figure}
    \centering
    \includegraphics[width=8cm,height=8cm,keepaspectratio]{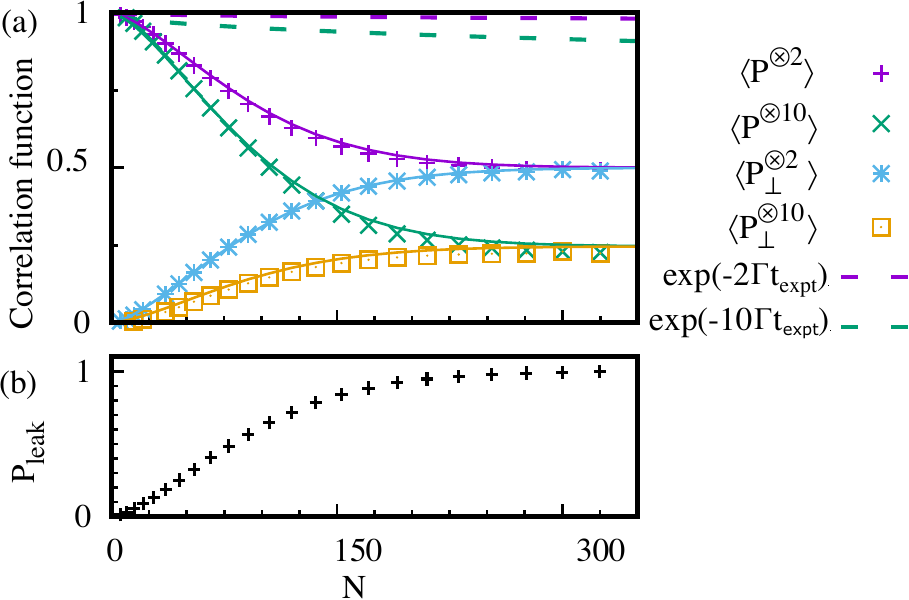}
    \caption{(a) Spin-spin correlation functions and (b) leakage probability for idealized GHZ state preparation with equal $J$, with $\delta = 2\pi \times 0.5$ kHz and other parameters from Fig.~\ref{fig:GHZ}(b). The correlation functions decay from their ideal values much faster than expected from the product of single-ion decoherences (dashed lines). Solid lines are Eqs.~(\ref{Pleak}) and (\ref{Pperpleak}).}
    \label{fig:correlation}
\end{figure}

Figure \ref{fig:correlation}(a) shows two-body and ten-body spin-spin correlation functions. There are significant deviations away from the ideal value $\langle P^{\otimes 2m}\rangle=1$, which exceed the decay $\exp(-2m \Gamma t_\text{expt})$ arising from the product of $2m$ single-ion decoherences (dotted lines), where $t_\text{expt}=2t_\text{arm}$ is the total length of the simulated experiment. The simple model below will attribute the additional error to dynamical interactions with leaked qubits; the probability for at least one qubit to leak is shown in Fig.~\ref{fig:correlation}(b).

If an ion leaks to $\ket{g}$, it generates an effective magnetic field $\exp(-i (\theta/2) \sum_{i\neq l} \sigma^z_i)$, where $\theta/2 \leq J t_\text{cat}/N = \pi/4$ for GHZ state preparation. In addition to this field, the remaining ions evolve under $\exp(-i (\pi/4) \sum_{i<j}^{N-1}\sigma_i^z\sigma^z_j)$, which alone would generate an $(N-1)$-ion GHZ state that is polarized in the spin direction $P_\perp$ perpendicular to the $N$-ion GHZ spin direction $P$ [where $P = \sigma^x$ when $N$ is even and $P = \sigma^y$ when $N$ is odd, see (\ref{GHZ})].  The effective magnetic field rotates the polarization phase $\theta$ of the $(N-1)$-ion GHZ state to $P_{\theta} = \cos(\theta)P_\perp \pm \sin(\theta) P$. In summary, single-ion leakage produces an $(N-1)-$ion GHZ state with a random phase, depending on the leakage time.  Multiple-ion leakage produces a smaller, but similarly randomized GHZ state.

Approximating the probability distribution over angles as uniform, we expect the projection along the intended spin orientation $P$ in the presence of leakage to be
\begin{align} \langle P^{\otimes 2m} \rangle_\text{leak} \approx \frac{2}{\pi} \int_0^{\pi/2} d\theta \bra{\psi_\text{GHZ}} P^{\otimes 2m}_{\theta} \ket{\psi_\text{GHZ}} \nonumber\\
= \frac{2}{\pi} \int_0^{\pi/2} d\theta \sin^{2m}(\theta) = \frac{1}{\sqrt{\pi}}\frac{\Gamma(\frac{1}{2}+m)}{\Gamma(1+m)} \end{align}
where the $\Gamma$ are Gamma functions. Using similar reasoning we obtain the same result for the perpendicular direction, $\langle P^{\otimes 2m}_\perp \rangle_\text{leak}=\langle P^{\otimes 2m} \rangle_\text{leak}$. 

We are now ready to estimate the $\langle P^{\otimes 2m}\rangle$ as a sum of parts with and without leakage, neglecting small errors from elastic and Raman scattering to simplify our reasoning.  In this approximation, the probability of ideal evolution is $P_\text{no\ leak} = \exp(-N\Gamma^{(0\to g)}t_\text{arm})$, with spin-correlation $\langle P^{\otimes 2m}\rangle_\text{no\ leak} = 1$, while the probability for leaky evolution is $P_\text{leak} = 1-P_\text{no\ leak}$, with spin-correlation $\langle P^{\otimes 2m}\rangle_\text{leak}$.  Then in total we expect
\begin{equation} \label{Pleak} \langle P^{\otimes 2m} \rangle \approx e^{-N\Gamma^{(0\to g)} t_\text{arm}} + \frac{(1-e^{-N\Gamma^{(0\to g)} t_\text{arm}})\Gamma(\frac{1}{2}+m)}{\sqrt{\pi}\Gamma(1+m)}. \end{equation}
We also expect that the perpendicular direction will have a nonzero correlation due solely to leakage
\begin{align} \label{Pperpleak} \langle P_\perp^{\otimes 2m} \rangle \approx  \frac{(1-e^{-N\Gamma^{(0\to g)} t_\text{arm}})\Gamma(\frac{1}{2}+m)}{\sqrt{\pi}\Gamma(1+m)}. \end{align}
The simple model (\ref{Pleak}-\ref{Pperpleak}) corresponds to the solid curves in Fig.~\ref{fig:correlation}(a). It accounts for the majority of the observed error in the exact results (data points) computed from relations equivalent to (\ref{main result})-(\ref{B}) but also including an analytic treatment of a spin-echo for $^{40}$Ca$^+$, as mentioned earlier and described in SI Sec.~C. To correct this error, postselection can be used to discard runs with leakage, with an expected sampling overhead of $1/(1-P_\text{leak})$. Repeated spin-echo sequences could mitigate the errors through partial cancellations. 

\begin{figure}
    \centering
    \includegraphics[height=6cm, width=7cm,keepaspectratio]{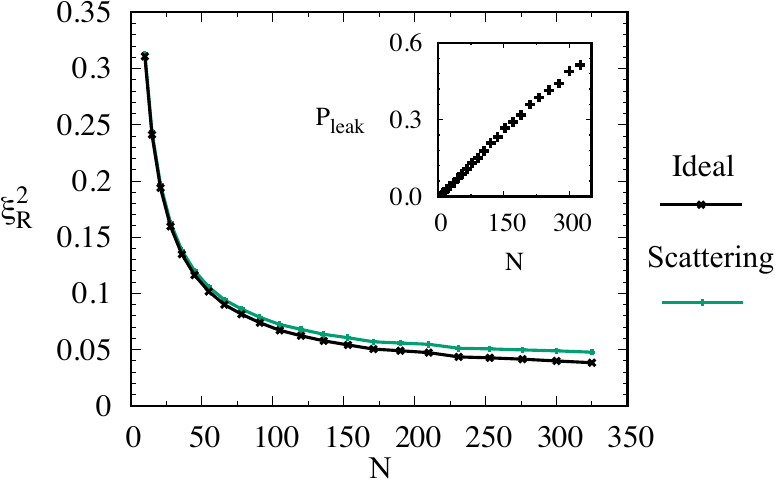}
    \caption{Optimal spin squeezing parameter $\xi^2_R$ with and without light scattering errors.  Inset: Leakage probability.} 
    \label{fig:spinsqueezing}
\end{figure}

\emph{Spin squeezing}.~Quantum metrology seeks to enhance sensing capabilities using quantum entanglement in protocols such as spin squeezing \cite{wineland1992spin,kitagawa1993squeezed,wineland1994squeezed,Bohnet_Science:2016,pezze2018quantum}. Squeezing  decreases the collective spin variance along a chosen axis to enhance sensitivity, as quantified by the spin squeezing parameter
 \begin{equation} \xi_R^2 = \min_\theta \frac{N(\Delta S^\theta)^2}{|\langle S^x\rangle|^2} \end{equation} 
 where $S^x = (1/2)\sum_i \sigma^x_i$ and $S^\theta = (1/2)\sum_i \cos(\theta)\sigma^z_i+ \sin(\theta)\sigma^y_i$ are collective spin vectors. In the unentangled initial state,
 $\xi_R^2=1$.  The Ising evolution (\ref{HIsing}) squeezes the state  through approximate ``one-axis twisting" leading to $\xi^2_R < 1$ \cite{pezze2018quantum}.
 
 Figure~\ref{fig:spinsqueezing} shows $\xi^2_R$ for a variety of sizes, with detuning $\delta = 2\pi \times 2$ kHz, $B$=0.9 T, and with the exact $J_{ij}$ for $^{40}$Ca$^+$.  In these results we minimized $\xi^2_R$ at each $N$ by varying the average coupling strengths $J/N$; best-fit results follow $Jt/N \approx 0.85N^{-0.62} \ll 1$ as shown in SI Sec.~E, which corresponds to a much weaker coupling than the one employed in GHZ state preparation (recall for GHZ states, $Jt_\text{cat}/N=\pi/4$).  We find that scattering causes only minor errors in $\xi^2_R$ relative to the ideal case, even with a few hundred ions where leakage is likely (Fig.~\ref{fig:spinsqueezing} inset). 

 We attribute the error-resilience of spin squeezing to weak coupling. If an ion leaks to its ground state then it generates an effective magnetic field $\exp(-i N^{-1}\sum_i J_{il}t_l\sigma^z_i)$ where $t_l$ is the time of leakage, but for weak-coupling $Jt/N \ll 1$ this field is close to the identity and does not contribute to significant error. Apart from this effective field, the ions also evolve under the $(N-1)$-ion version of $\Hc$, which leads to a $\xi^2_R$ similar to the case without leakage.  We find similar results for a simple benchmarking implementation of single-layer QAOA in SI Sec.~E.

\emph{Conclusions}.~We analytically solved a model of metastable-qubit Ising dynamics in the presence of light scattering errors, including inelastic, elastic, and leakage scattering channels, as pictured in Fig.~\ref{level diagram}(b). These analytic results allow for modelling of experiments with large numbers of ions $N$, where numerical simulation of the quantum state is prohibitively difficult or impossible. 

Our main findings with respect to the physical effect of leakage are as follows: 1) When a qubit leaks outside the qubit manifold, then its Ising interaction with the other qubits behaves as an ``effective magnetic field" that is applied up until the leakage event occurs.  The strength of the field depends on the coupling strength and on the specific leakage time, the latter of which is random. The physical effect of these fields is that all nonleaked qubits will rotate between the Pauli spin directions $\sigma^x$ and $\sigma^y$. 2) 
Depending on the coupling-time product $Jt/N$ that is present in a given experiment, the effective magnetic fields that are generated in leakage may have catastrophic or negligible effects. 
When $Jt/N \approx 1$, as in GHZ state preparation or strongly-coupled quantum simulations, then the effective magnetic fields completely randomize the Pauli spin directions $\sigma^x$ and $\sigma^y$ for all nonleaked qubits. When leakage occurs, these errors must be addressed to obtain accurate results, for example, by postselecting on the presence of even a single leaked qubit.  When $Jt/N \ll 1$, as in spin-squeezing, then the nonleaked qubits rotate by only a negligible angle under the effective magnetic fields. Therefore, in this case leakage does not significantly affect accuracy. 

In conclusion, we have developed analytic theory that quantifies and provides physical insights into leakage errors in trapped-ion quantum simulations. In this context,  observables in a given experiment can be readily computed using the relevant Ising couplings, scattering rates, and experimental runtimes. We expect our results to be useful in estimating experimental tradeoffs obtained for different choices of experimental parameters, such as coupling strengths and sampling overheads for post selection, as well as modelling a variety of future experiments. 

Code verifying the correctness of the analytical results is available at \href{https://code.ornl.gov/5ci/exactly-solved-model-of-light-scattering-errors-in-quantum-simulations-with-metastable-trapped-ion-qubits}{https://code.ornl.gov/5ci/exactly-solved-model-of-light-scattering-errors-in-quantum-simulations-with-metastable-trapped-ion-qubits}

\begin{acknowledgments}
The authors would like to thank John Bollinger, Joseph Wang, Bryan Gard, and Ryan Bennink for providing useful comments on this manuscript. 
This material is based upon work supported by the Defense Advanced Research Projects Agency (DARPA) under Contract No. HR001120C0046.
\end{acknowledgments}

\bibliographystyle{unsrt}
\bibliography{refs}

\begin{widetext}

\section*{Supplemental Material}

\appendix

\section{Summary of theoretical results} \label{results summary}

We have derived exact expectation values for spin operators, for dynamics under the master equation (4) in the main text, and for an initial state $\ket{\psi_0}=\ket{+}^{\otimes N}$. There are a variety of different types of terms, corresponding to distinct types of quantum-state trajectories as summarized in Table \ref{tab:trajectory_terms}; the expressions are presented below and make use of several derived parameters summarized in Table \ref{derived terms}. 

First we consider operators that completely characterize a single-qubit reduced state.  These include populations described by $ \sigma^z_j = \ket{z_j}\bra{z_j}$ where $z \in \{0,1,g\}$, as well as raising and lowering operators  $\sigma^+_j = \ket{0_j}\bra{1_j}$ and $\sigma^-_j = \ket{1_j}\bra{0_j}$ describing coherences in the qubit manifold; note that  $\sigma^+_j$ and $\sigma^-_j$ raise and lower the energy, not the index $0/1$ of the state, see Fig.~1 in the main paper.  There is no coherence between $\ket{g}$ and the qubit levels $\ket{0},\ket{1}$, so operators such as $\ket{g}\bra{0}$ have zero expectation, and we ignore them in what follows. 

Expectation values of these operators on a qubit  $j$ correspond to individual matrix elements in the reduced density matrix $\rho_j = \mathrm{Tr}_{i \neq j} \rho$, since for example $\langle \sigma^+_j \rangle = \mathrm{Tr}\rho_j \sigma^+_j = \bra{1_j} \rho_j \ket{0_j}$ which is a single matrix element of $\rho_j$; the diagonal elements are given similarly by the $\langle \sigma_j^z\rangle$; together these operators give a complete analytic description of the reduced state $\rho_j$.  The mathematical expressions are (with $\nu \in \{+,-\})$
\begin{align} \label{1 qubit RDM} \langle \sigma_j^\nu \rangle & = \frac{e^{-\Gamma t}}{2} \prod_{i\neq j} [I(\nu J_{ij},t) + R(\nu J_{ij},t) + L(\nu J_{ij},t) + B(\nu J_{ij},t)], \nonumber\\
\langle \sigma_j^0 \rangle & = I^{(0)}(0,t) + R^{(0)}(0,t), \nonumber\\
\langle \sigma_j^1 \rangle & = I^{(1)}(0,t) + R^{(1)}(0,t), \nonumber\\
\langle \sigma_j^g \rangle & = L(0,t) + B(0,t)
\end{align} 
where $\Gamma = \sum_\alpha \Gamma^{(\alpha)}/2$ is the single-qubit decoherence rate, with the notation $\nu J_{ij} = J_{ij}$ for $\nu=+$ and $\nu J_{ij}=-J_{ij}$ for $\nu=-$. These relations can be used to compute the expectation of any single-qubit operator, for example, the Pauli operator expectation values are
\begin{align} 
    \langle \sigma^x_j\rangle = \langle \sigma_j^+\rangle + \langle \sigma_j^-\rangle, \nonumber\\
    \langle \sigma^y_j\rangle = -i\langle \sigma_j^+\rangle + i\langle \sigma_j^-\rangle, \nonumber\\
    \langle \sigma^z_j\rangle = \langle \sigma_j^0\rangle - \langle \sigma_j^1\rangle
\end{align}
We now describe the functions $I,R,L$ and $B$, followed by the extension to arbitrary $m$-body observables.

\begin{table}[]
    \centering
    \caption{Summary of the trajectories that produce each term in \ref{1 qubit RDM}.}
    \begin{tabular}{|c|c|}
    \hline
    $I(J_{ij},t)$ & Ideal evolution  \\
    \hline
    $R(J_{ij},t)$ & Raman transitions in the qubit manifold\\
    \hline
    $L(J_{ij},t)$ & Leakage \\        
    \hline
    $B(J_{ij},t)$ & Both Raman and Leakage \\        
    \hline
    \end{tabular}
    \label{tab:trajectory_terms}
\end{table}

\begin{table}
\caption{Mathematical quantities used in the results and their physical meanings.}
\label{derived terms}
\begin{tabular}{|c|c|}
\hline
$\Gamma = \sum_\alpha \Gamma^{(\alpha)}/2$ & Single-qubit decoherence rate\\
\hline
$\Gamma^{(L)} = (\Gamma^{(0 \to g)} + \Gamma^{(1 \to g)})/2 $ & Single-qubit leakage scattering decoherence rate\\
\hline
$\Gamma^{(R)} = (\Gamma^{(0 \to 1)} + \Gamma^{(1 \to 0)})/2 $ & Single-qubit Raman scattering decoherence rate\\
\hline
$\Gamma^{(B)} = (\Gamma^{(0 \to 1)}\Gamma^{(1\to g)} + \Gamma^{(1 \to 0)}\Gamma^{(0\to g)})/2 $ & Combined rate for leakage and Raman scattering\\
\hline
$\Gamma^{(0)} = \Gamma^{(0 \to 1)} + \Gamma^{(0 \to g)} $ & Single-qubit scattering rate from $\ket{0}$\\
\hline
$\Gamma^{(1)} = \Gamma^{(1 \to 0)} + \Gamma^{(1 \to g)} $ & Single-qubit scattering rate from $\ket{1}$\\
\hline
$\lambda = (\Gamma^{(0)} + \Gamma^{(1)})/2$ & Single qubit decoherence rate due to inelastic scattering \\
\hline
$\Delta = (\Gamma^{(0)} - \Gamma^{(1)})/2$ & Decoherence rate difference \\
\hline
$\Delta^{(L)} = (\Gamma^{(0\to g)} - \Gamma^{(1\to g)})/2$ & Leakage decoherence rate difference \\
\hline
$s(J_{ij}) = 2J_{ij}/N + i\Delta$ & Damped oscillation frequency of $\Hc_\text{eff}$ \\
\hline
$\zeta(J_{ij}) = \sqrt{s^2 - \Gamma^{(1\to0)}\Gamma^{(0\to1)}}$ & Damped oscillation frequency in sequences of Raman transitions \\
\hline
\end{tabular}
\end{table}

The term
\begin{align} \label{IJij} I(\nu J_{ij},t) = e^{- \lambda t} \cos(s(\nu J_{ij})t). \end{align}
corresponds to ideal evolution. The exponent relates to the single-qubit decoherence rate due to inelastic scattering $\lambda = (\Gamma^{(0 \to 1)} + \Gamma^{(0 \to g)} + \Gamma^{(1 \to 0)} + \Gamma^{(1 \to g)})/2$, while the complex number $s(\nu J_{ij}) = 2\nu J_{ij}/N + i\Delta$ describes the  oscillations generated under $\Hc_\text{eff}$, including both the signed Ising coupling $\nu J_{ij}$ as well as the elastic scattering decoherence rate asymmetry $\Delta =(\Gamma^{(0 \to 1)} + \Gamma^{(0 \to g)} - \Gamma^{(1 \to 0)} - \Gamma^{(1 \to g)})/2 $. The total term $I(\nu J_{ij},t)$ can also be separated into components $I(\nu J_{ij},t) = I^{(0)}(\nu J_{ij},t) + I^{(1)}(\nu J_{ij},t)$ related to the contributions from individual basis states $\ket{0_i}$ and $\ket{1_i}$, 
\begin{equation} I^{(0)}(\nu J_{ij},t) = \frac{e^{i2\nu J_{ij}t/N}e^{-\Gamma^{(0)}t}}{2}, \ \ I^{(1)}(\nu J_{ij},t) = \frac{e^{-i2\nu J_{ij}t/N}e^{-\Gamma^{(1)}t}}{2}, \end{equation} 
where $\Gamma^{(0)} = \Gamma^{(0\to1)} + \Gamma^{(0 \to g)}$ describes the total decay rate from $\ket{0_i}$, and similarly for $\Gamma^{(1)}$. 

The second term $R(\nu J_{ij},t)$ corresponds to trajectories under sequences of Raman transitions that do not leave the qubit manifold,
\begin{equation} R(\nu J_{ij},t) = e^{-\lambda t}\left[\cos(\zeta(\nu J_{ij}) t) - \cos(s(\nu J_{ij})t) + \Gamma^{(R)}t\mathrm{sinc}(\zeta(\nu J_{ij}) t) \right],\end{equation}
 where $\Gamma^{(R)}=(\Gamma^{(0\to 1)} + \Gamma^{(1\to 0)})/2$ is the single-qubit Raman decoherence rate and the complex number $\zeta(\nu J_{ij}) = \sqrt{s^2(\nu J_{ij}) - \Gamma^{(0 \to 1)}\Gamma^{(1\to 0)}}$ is the effective damped frequency for dynamics with Raman transitions within the qubit manifold.  Similar to the previous case, the $R$ terms can also be separated into components for which the trajectory terminates in either state $\ket{0_i}$ or $\ket{1_i}$, with $R(\nu J_{ij},t) = R^{(0)}(\nu J_{ij},t) + R^{(1)}(\nu J_{ij},t)$ and
\begin{align} R^{(0)}(\nu J_{ij},t) = \frac{e^{-\lambda t}}{2} \left[\cos(\zeta t) - \cos(st) + \Gamma^{(1\to0)} t\mathrm{sinc}(\zeta t) + ist\left( \mathrm{sinc}(\zeta t) - \mathrm{sinc}(st)\right) \right], \nonumber\\
R^{(1)}(\nu J_{ij},t) = \frac{e^{-\lambda t}}{2} \left[\cos(\zeta t) - \cos(st) + \Gamma^{(0\to1)}t \mathrm{sinc}(\zeta t) - ist\left(\mathrm{sinc}(\zeta t) - \mathrm{sinc}(st)\right) \right] \label{R01} \end{align}
where we have removed the arguments in $s=s(\nu J_{ij})$ and $\zeta = \zeta(\nu J_{ij})$ to shorten the notation. 

Both $I(\nu J_{ij},t)$ and $ R(\nu J_{ij,t})$ were first derived by Foss-Feig {\it et. al} \cite{Foss-Feig_PRA:2013,fossfeigdissertation}, with the minor variation here that decays from all jump operators enter into $\lambda$ and $\Delta$.  In the notation of Eq.~(10) in Ref.~\cite{Foss-Feig_PRA:2013}, when $\Gamma^{(0\to g)}=\Gamma^{(1\to g)}=0$ we have $I(\nu J_{ij},t) + R(\nu J_{ij,t}) = \Phi(\nu J_{ij},t)$. 

The final two terms describe leakage outside of the qubit manifold; $L(\nu J_{ij},t)$ describes leakage without prior Raman transitions while $B(\nu J_{ij},t)$ describes trajectories with both a sequence of Raman transitions within the qubit manifold and a final leakage transition.  These terms are
\begin{align} L(\nu J_{ij},t) = \frac{\Gamma^{(0\to g)} f(\Gamma^{(0)},2\nu J_{ij}/N,t) + \Gamma^{(1\to g)} f(\Gamma^{(1)},-2\nu J_{ij}/N,t)}{2}\end{align}
where we have defined 
\begin{align} f(\Gamma,\chi,t)= \int_0^t dt' e^{(i\chi - \Gamma)t'} =  e^{ i(\chi + i\Gamma)t/2} t\mathrm{sinc}[(\chi + i\Gamma)t/2] \end{align}
and
\begin{align} B(\nu J_{ij},t) & = \Gamma^{(L)}[a(\lambda,\zeta,t) - a(\lambda,s,t)] + \Gamma^{(B)}b(\lambda,\zeta,t) + is\Delta^{(L)}[b(\lambda,\zeta,t)-b(\lambda,s,t)] \end{align}
where $\Delta^{(L)} = (\Gamma^{(0\to g)}-\Gamma^{(1\to g)})/2$,  $\Gamma^{(B)}=(\Gamma^{(0\to 1)}\Gamma^{(1\to g)} + \Gamma^{(1\to 0)}\Gamma^{(0\to g)})/2,$ and the functions
\begin{align} a(\Gamma,\chi,t) = \int_0^t dt' e^{-\Gamma t'} \cos(\chi t) =  \frac{f(\Gamma,\chi,t) + f(\Gamma,-\chi,t)}{2} \nonumber\\
b(\Gamma,\chi,t) = \int_0^t dt' e^{-\Gamma t'} t\mathrm{sinc}(\chi t) =  -\frac{i}{\chi}\frac{f(\Gamma,\chi,t) - f(\Gamma,-\chi,t)}{2} \end{align}

We now consider expectations of arbitrary $m$-body operators, computed from the generic operator product
\begin{equation} \bm \sigma = \prod_{a \in P^{(0)}} \sigma_a^0 \prod_{b \in P^{(1)}} \sigma_b^1 \prod_{c \in P^{(g)}} \sigma_c^g \prod_{j \in M} \sigma_j^{\nu_j} \end{equation}
where $M = \{q_1, \ldots, q_m\}$ is the set of $m$ qubits with raising and lowering operators in the product, while $P^{(0)}$, $P^{(1)}$, and $P^{(g)}$ are the respective sets of qubits with projectors $\sigma^0, \sigma^1,$ and $\sigma^g$ in the product.
We derived its expectation as
\begin{align} \label{generic expectation value} \left\langle \bm \sigma \right\rangle = &\frac{e^{-m\Gamma t}}{2^m} F^{(0)}(\bm \nu, \bm J,t) F^{(1)}(\bm \nu, \bm J,t) F^{(g)}(\bm \nu, \bm J,t)F(\bm \nu, \bm J,t) \end{align}
with
\begin{align} F^{(0)}(\bm \nu, \bm J,t) & = \prod_{a \in P^{(0)}} \left[I^{(0)}\left(\sum_{j \in M}\nu_j J_{aj},t\right) + R^{(0)}\left(\sum_{j \in M}\nu_j J_{aj},t\right)\right] \nonumber\\
F^{(1)}(\bm \nu, \bm J,t) & = \prod_{b \in P^{(1)}} \left[I^{(1)}\left(\sum_{j \in M}\nu_j J_{bj},t\right) + R^{(1)}\left(\sum_{j \in M}\nu_j J_{bj},t\right)\right] \nonumber\\
F^{(g)}(\bm \nu, \bm J,t) & = \prod_{c \in P^{(g)}} \left[L\left(\sum_{j \in M}\nu_j J_{cj},t\right) + B\left(\sum_{j \in M}\nu_j J_{cj},t\right)\right] \nonumber\\
F(\bm \nu, \bm J,t) & = \prod_{i\notin M,P^{(0)},P^{(1)},P^{(g)}} \left[I\left(\sum_{j \in M}\nu_j J_{ij},t\right) + R\left(\sum_{j \in M}\nu_j J_{ij},t\right) + L\left(\sum_{j \in M}\nu_j J_{ij},t\right) + B\left(\sum_{j \in M}\nu_j J_{ij},t\right)\right] \label{F functions}\end{align}

Using equations (\ref{generic expectation value}-\ref{F functions}) the expected value $\langle O \rangle$ of any operator $O$ can be computed. To do this,  we first express $O$ in the basis $\ket{\bm z} = \ket{z_1,z_2,\ldots,z_N}$ where $z_i \in \{0,1,g\}$ is the basis element of the $i$th qubit. An operator $O$ can be expanded in terms of this basis as 
\begin{equation} O =  \sum_{\bm z, \bm z'} \ket{\bm z}\bra{\bm z}O\ket{\bm z'}\bra{\bm z'} =   \sum_{\bm z, \bm z'} O_{\bm z, \bm z'}\ket{\bm z}\bra{\bm z'}\end{equation} 
where $O_{\bm z, \bm z'}$ is a matrix element of $O$. The outer product $\ket{\bm z}\bra{\bm z'}$ on the right can be expressed as products of the $\sigma^0, \sigma^1,$ $\sigma^g$, $\sigma^+$, and $\sigma^-$: $\ket{\bm z}\bra{\bm z'} = \bigotimes_i \ket{z_i}\bra{z_i'} = \bigotimes_i \sigma^{\alpha_i}_i = \bm \sigma$, where $\sigma^{\alpha_i}_i = \ket{z_i} \bra{z_i'}$.  
 This ignores terms containing products such as $\ket{0_i}\bra{g_i}$ describing coherence between $\ket{g_i}$ and the qubit levels, since the expectation of these terms is zero in our model, as mentioned previously. Restricting $O$ to the relevant basis spanned by the $\bm \sigma$,  we have $O = \sum_{\bm \sigma} O_{\bm \sigma} \bm \sigma$, where $O_{\bm \sigma} = \mathrm{Tr} (\bm \sigma^\dag O)$. Then by definition 
 \begin{equation} \langle O \rangle = \mathrm{Tr}(\rho O) = \sum_{\bm \sigma} O_{\bm \sigma} \langle \bm \sigma \rangle \end{equation}
where $\langle \bm \sigma\rangle$ is computed from (\ref{generic expectation value}).

The entire quantum state $\rho$ can be computed using the standard operator expansion
\begin{equation}\rho = \sum_{\bm \sigma} \bm \sigma^\dag \frac{\mathrm{Tr}(\rho \bm \sigma)}{\sqrt{\mathrm{Tr}(\bm \sigma^\dag \bm \sigma)}} = \sum_{\bm \sigma} \bm \sigma^\dag \langle \bm \sigma \rangle\end{equation}
while reduced density operators are computed similarly, but with a sum over $\bm \sigma$ that span the relevant Liouville subspace of the reduced set of qubits. 

\section{Derivation of the main results}

Here we present the derivation of Eq.~(\ref{generic expectation value}).  We begin with an overview of the analytic ``quantum trajectories" approach, then derive the individual terms $I,R,L,B$ appearing in (\ref{generic expectation value}).  In each case we shall focus on evaluating $\langle \sigma^+_j\rangle$; the more general results for arbitrary operator products follow almost immediately afterwards, as we describe towards the end of Sec.~\ref{leakage appendix}.

\subsection{Overview of quantum trajectories approach} 
The quantum trajectories approach is a systematic way to organize terms that are generated by continuous evolution under the master equation. The time evolution of an initial pure state can be viewed as generating an ensemble of pure state trajectories that can be reasoned about analytically. For example, given an initial pure state $\rho(0) = \ket{\psi}\bra{\psi}$, a differential timestep $\delta t$ produces a state $\rho(\delta t) = \rho(0) + \dot \rho(0) \delta t$ at leading order.  From the master equation
\begin{align} \rho(\delta t) = \rho(0) -i\delta t(\Hc_\text{eff}\rho(0)-\rho(0) \Hc_\text{eff}^\dag) + 2\delta t\sum_{j=1}^N \sum_{\alpha \in A} \Jc_j^{(\alpha)} \rho(0) {\Jc_j^{(\alpha)}}^\dag \end{align}
The terms $\rho(0) -i\delta t(\Hc_\text{eff}\rho(0)-\rho(0) \Hc_\text{eff}^\dag)$ represent Schr\"odingerlike evolution under the effective Hamiltonian $\Hc_\text{eff}$.  The last term represents a variety of different quantum jump processes.  Importantly, both the ideal evolution and the individual jump components $\Jc_j^{(\alpha)}\rho(0){\Jc_j^{(\alpha)}}^\dag $ are unnormalized pure states---rather than dealing with the mixed total state $\rho(\delta t)$ on the left, we can equivalently consider the ensemble of $1+N|A|$ pure-state trajectories on the right, where $|A|$ is the number of different types of jump operators, each of which can act on any one of $N$ qubits. The norm of each state in the ensembles relates to its probability, as described below. 

Further propagating the time evolution yields a state $\rho(2\delta t)$, in which each pure state on the right is propogated into $1+N|A|$ new pure states. Repeating this process leads to an exponentially large number of states, but it can be simplified by noting that many of them share similar structures, and that many of the states have zero norm (hence zero probability) because repeated applications of a jump operators such as $\Jc_j^{(0\to1)}$ ``deletes" a trajectory since  ${\Jc_j^{(0\to1)}}^2 \sim \ket{1}\bra{0}1\rangle\bra{0} = 0$. 

A single quantum-state trajectory $\ket{\overline{\psi}(t;T)}$ at time $t$ is specified by a set of quantum jump operators $J^{(\alpha)}_i$ and associated jump times $t_i$, as denoted in the trajectory parameter $T = \{(\Jc^{(\alpha_s)}_s, t_s) : s \in S\}$ where $S$ is the set of all quantum jump events. The unitary dynamics of a trajectory with $m$ jumps has the mathematical form 
\begin{equation} \label{trajectory unitary} \overline{U}(t;T) = e^{-i\Hc_\mathrm{eff}t-t_m} \sqrt{2}\Jc_{j_m}^{(\alpha_m)} e^{-i\Hc_\mathrm{eff}(t_m-t_{m-1})} \sqrt{2}\Jc_{j_{m-1}}^{(\alpha_{m-1})} e^{-i\Hc_\mathrm{eff}(t_{m-1}-t_{m-2})} \ldots \sqrt{2}\Jc_{j_1}^{(\alpha_{1})} e^{-i\Hc_\mathrm{eff}t_1} \end{equation} 
and this produces the nonnormalized trajectory state
\begin{equation} \label{trajectory state} \ket{\overline{\psi}(t;T)} =\overline U(t;T) \ket{\psi(0)} \end{equation}
where the overlines on $\overline{U}$ and $\overline{\psi}$ signify the dynamics and final state are not normalized, respectively. The factors of $\sqrt{2}$ in the trajectory come from the definition of $\Dc(\rho)$. 

The total density matrix evolution comes from the sum over all possible nonnormalized trajectories, that is, all terms obtained in a sequence of differential timesteps following the master equation,
\begin{equation} \label{rho sum unnormalized} \rho(t) = \sum_T \ket{\overline{\psi}(t;T)}\bra{\overline{\psi}(t;T)} .\end{equation} 
It is also equivalent to consider $\rho(t)$ as a probabilistic ensemble over normalized states 
\begin{equation} \ket{\psi(t;T)} = \frac{\ket{\overline{\psi}(t;T)}}{\sqrt{\langle \overline{\psi}(t;T) \ket{\overline{\psi}(t;T)}}},\end{equation}
as in Ref.~\cite{Foss-Feig_PRA:2013}, with the density operator expressed as
\begin{equation} \rho(t) = \sum_T \mathrm{Pr}(T)\ket{\psi(t;T)}\bra{\psi(t;T)}, \end{equation} 
with $\mathrm{Pr}(T) = \langle \overline{\psi}(t;T)\ket{\overline{\psi}(t;T)}$ the probability of trajectory $T$.

\subsection{Leakage trajectories} \label{leakage appendix}

It will be useful to first focus on trajectories that only include the jump operators $\Jc^{(0\to g)}$ and $\Jc^{(1\to g)}$ in the trajectory parameter $T_L = \{(\Jc_{k_0}^{(0\to g)},t_{k_0}), (\Jc_{k_1}^{(1\to g)},t_{k_1}): k_0 \in K_0, k_1 \in K_1\}$, where $K_0$ and $K_1$ are the respective sets of qubits undergoing jumps $J_{k_0}^{(0\to g)}$ and $J_{k_1}^{(1\to g)}$ at times $t_{k_0}$ and $t_{k_1}$. Analyzing these trajectories will provide a straightforward demonstration of how to derive operator expectation values from the quantum trajectories approach and compute some necessary terms for the final result, which will include additional trajectories that we treat in later sections.

A trajectory with nonzero norm can have at most one of $\Jc^{(0\to g)}$ or $\Jc^{(1\to g)}$ applied to each qubit, since applying either of these operators takes a qubit into the level $\ket{g}$, while an additional application of an operator such as  $\Jc^{(1\to g)} \sim \ket{g}\bra{1}$ will result in an inner product $\bra{1}g\rangle=0$ when applied to a state already in $\ket{g}$. A generic trajectory containing ideal evolution along with $\Jc^{(0\to g)}$ and $\Jc^{(1\to g)}$ jump operators will produce a state [following from the dynamics Eqs.~(\ref{trajectory unitary})-(\ref{trajectory state})]
\begin{align} \label{leakage trajectory} \ket{\overline{\psi}(t;T_L)} = e^{-i \Hc(T_L)t} & \prod_{k_0 \in K_0} \frac{\sqrt{\Gamma^{(0\to g)} (\delta t)_{k_0}} e^{-t_{k_0} \Gamma^{(0)} /2} \ket{g_{k_0}}}{\sqrt{2}} \prod_{k_1 \in K_1} \frac{\sqrt{\Gamma^{(1\to g)} (\delta t)_{k_1}} e^{-t_{k_1} \Gamma^{(1)} /2} \ket{g_{k_1}}}{\sqrt{2}}
\nonumber\\
& \times \prod_{i \notin K} \frac{e^{-\Gamma^{(0)}t/2}\ket{0_i} + e^{-\Gamma^{(1)}t/2}\ket{1_i}}{\sqrt{2}} \end{align}
where $K = K_0\cup K_1$ is the total set of qubits that decay to $\ket{g}$. The trajectory-dependent Hamiltonian is
\begin{equation} \label{HT leakage} \Hc(T_L) = \frac{1}{N}\sum_{i,j \notin K_0, K_1} J_{ij} \sigma^z_i \sigma^z_j + \frac{1}{N}\sum_{k_0 \in K_0} (t_{k_0}/t) \sum_{i \notin K} J_{ik_0} \sigma^z_i - \frac{1}{N}\sum_{k_1 \in K_1} (t_{k_1}/t) \sum_{i \notin K} J_{ik_1} \sigma^z_i \end{equation}
The form of this Hamiltonian follows from reasoning originally described in Ref.~\cite{Foss-Feig_PRA:2013}; a jump operator such as $\Jc^{(0\to g)}_{k_0} \sim \ket{g_{k_0}}\bra{0_{k_0}}$ will remove the component of the quantum state for which the qubit was in $\ket{1_{k_0}}$ prior to the jump, so the final state will be identical to the state that would be produced if the initial state of the effected qubit had simply been $\ket{0_{k_0}}$.  Since $\ket{0_{k_0}}$ is an eigenstate of $\sigma^z_{k_0}$, we can replace the Pauli operator with its eigenvalue $\sigma^z_{k_0} \to 1$ in the Hamiltonian $\Hc(T_L)$.  This yields a set of ``effective magnetic field" terms $N^{-1}\sum_{i \notin K} J_{ik_0} \sigma_i^z$ in place of the original Ising couplings associated with $k_0$, and similarly for the $k_1$ terms with $\sigma^z_{k_1} \to -1.$ Together these types of jumps yield the effective magnetic field terms on the right (2nd and 3rd terms). These fields are active until the times $t_{k_i}$, at which point the qubit decays to $\ket{g_{k_i}}$; the coefficients $(t_{k_i}/t)$ ensure the fields are applied for the correct times $t_{k_i}$ in the Hamiltonian evolution $\exp(-i \Hc(T_L)t)$.  We do not consider Ising terms for which both qubits decay to the ground state, because these only contribute physically-irrelevant global phases. 

We now compute the trajectory-dependent expectation value $\bra{\overline{\psi}(t;T_L)}\sigma_j^+ \ket{\overline{\psi}(t;T_L)}$ for the generic leakage trajectory in (\ref{leakage trajectory}), which constitutes a building block for more complex terms introduced later. For trajectories where qubit $j$ jumps to the ground state, the expectation $\bra{\overline{\psi}(t;T_L)}\sigma_j^+ \ket{\overline{\psi}(t;T_L)}=0$ since $\sigma_j^+\ket{g}=0$. To analyze the case where qubit $j$ does not jump we begin with the Heisenberg representation of $\sigma_j^+$,
\begin{equation} \label{sigma plus Heisenberg} e^{i \Hc(T_L)t} \sigma_j^+ e^{- i \Hc(T_L)t} = \sigma_j^+ e^{i2tN^{-1}\sum_{i \notin K} J_{ij}\sigma^z_i}e^{i2N^{-1}\sum_{k_0 \in K_0} t_{k_0}J_{k_0j}}e^{-i2N^{-1}\sum_{k_1 \in K_1} t_{k_1}J_{k_1j}} \end{equation} 
then compute the expectation as
\begin{align} \bra{\overline{\psi}(t;T_L)}\sigma_j^+ \ket{\overline{\psi}(t;T_L)}   = & \frac{e^{-\lambda t}}{2}\prod_{k_0 \in K_0} \frac{\Gamma^{(0\to g)} (\delta t)_{k_0} e^{it_{k_0} (2J_{k_0j}/N+i\Gamma^{(0)}) }}{2} \prod_{k_1 \in K_1} \frac{\Gamma^{(1\to g)} (\delta t)_{k_1} e^{-it_{k_1} (2J_{k_1j}/N-i\Gamma^{(0)}) }}{2} \nonumber\\ 
& \times \prod_{i \notin K} \left(I^{(0)}(J_{ij},t) + I^{(1)}(J_{ij},t) \right) \end{align}
where $\lambda = (\Gamma^{(0)} + \Gamma^{(1)})/2$. The functions 
\begin{equation} \label{I01} I^{(0)}(J_{ij},t) = \frac{e^{i2tJ_{ij}/N}e^{-\Gamma^{(0)}t}}{2}, \ \ I^{(1)}(J_{ij},t) = \frac{e^{-i2tJ_{ij}/N}e^{-\Gamma^{(1)}t}}{2}, \end{equation} 
respectively describe the contribution from populations in $\ket{0_i}$ and $\ket{1_i}$ for qubits that undergo ideal evolution (without leakage).  We denote the sum of these terms
\begin{equation} I(J_{ij},t)=I^{(0)}(J_{ij},t)+I^{(1)}(J_{ij},t) = e^{-\lambda t}\cos(s(J_{ij})t) \end{equation}
where $s = 2J_{ij}/N + i\Delta$ is the effective damped oscillation frequency, including the influence of asymmetric inelastic scattering through $\Delta = (\Gamma^{(0)}-\Gamma^{(1)})/2.$

For the given sets of leaked qubits $K_0$ and $K_1$ we now compute the expectation $\langle \sigma_j^+(K_0,K_1)\rangle$ including contributions from all leakage times.  We will then sum $\langle \sigma_j^+(K_0,K_1)\rangle$ over all $K_0, K_1$ to obtain the total $\langle \sigma_j^+\rangle$ as a sum over all possible trajectories $T_L$. 

To account for  all possible leakage times we need to sum over all possible choices of $t_{k_{i}}$ in the limit $\delta t \to 0$, which produces Reimann integrals
\begin{align} \langle \sigma_j^+(K_0,K_1)\rangle & = \lim_{\delta t \to 0} \sum_{t_{k_0} ; k_0 \in K_0} \sum_{t_{k_1} ; k_1 \in K_1} \bra{\overline{\psi}(t;T_L)}\sigma_j^+ \ket{\overline{\psi}(t;T_L)} \nonumber\\ 
& = \frac{e^ {-\lambda t}}{2}\prod_{k_0 \in K_0} \frac{\Gamma^{(0\to g)} \int_0^t dt_{k_0} e^{it_{k_0} (2J_{k_0j}/N+i\Gamma^{(0)}) }}{2} 
 \prod_{k_1 \in K_1} \frac{\Gamma^{(1\to g)} \int_0^tdt_{k_1} e^{-it_{k_1} (2J_{k_1j}/N-i\Gamma^{(1)}) }}{2} \prod_{i \notin K} I(J_{ij},t) \nonumber\\
 & = \frac{e^ {-\lambda t}}{2}\prod_{k_0 \in K_0} \frac{\Gamma^{(0\to g)} f(\Gamma^{(0)},2J_{k_0j}/N,t)}{2}
 \prod_{k_1 \in K_1} \frac{\Gamma^{(1\to g)} f(\Gamma^{(1)},-2J_{k_1j}/N,t)}{2}\prod_{i \notin K} I(J_{ij},t) \nonumber\\
 & = \frac{e^ {-\lambda t}}{2} \prod_{k_0 \in K_0} L^{(0 \to g)}(J_{k_0j},t) \prod_{k_1 \in K_1} L^{(1 \to g)}(J_{k_1j},t) \prod_{i \notin K} I(J_{ij},t) \label{leak trajectory}\end{align}
where 
\begin{align} f(\Gamma,\chi,t)= \int_0^t dt' e^{(i\chi - \Gamma)t'} =  e^{ i(\chi + i\Gamma)t/2} t\mathrm{sinc}[(\chi + i\Gamma)t/2] \end{align}
and the functions
\begin{align}\label{L01} L^{(0 \to g)}(J_{k_0j},t) = \frac{\Gamma^{(0\to g)} f(\Gamma^{(0)},2J_{k_0j}/N,t)}{2},\nonumber\\
L^{(1 \to g)}(J_{k_1j},t)= \frac{\Gamma^{(1\to g)} f(\Gamma^{(1)},-2J_{k_1j}/N,t)}{2} \end{align}
describe contributions from qubits that leaked from $\ket{0}$ and $\ket{1}$ respectively. 

We now consider the total expectation value for $\langle \sigma_j^+\rangle$, which is computed from the total set of all trajectories (\ref{rho sum unnormalized}) including the ideal trajectory plus all leakage trajectories of the type described above.  Recall that only trajectories where qubit $j$ does not leak will contribute to the sum, since $\sigma^+_j\ket{g_j}=0$. For the other qubits we need to sum over all combinations of ideal evolution, jumping $\ket{0_i} \to \ket{g_i}$, or jumping $\ket{1_i} \to \ket{g_i}$. To compute this we sum (\ref{leak trajectory}) over all sets of leaked qubits $K_0$ and $K_1$ 
\begin{align} \label{sigma plus leakage final} \langle \sigma_j^+ \rangle & = \sum_{T_L}\bra{\overline{\psi}(t;T_L)}\sigma_j^+ \ket{\overline{\psi}(t;T_L)}  =   \sum_{K_0,K_1} \langle \sigma_j^+(K_0,K_1)\rangle =  \frac{e^{-\lambda t}}{2} \prod_{i\neq j} \left[I(J_{ij},t)  + L(J_{ij},t) \right] \end{align}
where 
\begin{equation} L(J_{ij},t) = L^{(0 \to g)}(J_{ij,t}) + L^{(1 \to g)}(J_{ij,t}) \end{equation}
Equation (\ref{sigma plus leakage final}) describes $\langle \sigma^+_j\rangle$ as a sum over all possible trajectories, including all combinations of ideal [$I(J_{ij},t)$] and leaky [$L(J_{ij},t)$] dynamics on each qubit.  This equates to our general result (\ref{generic expectation value}) in the absence of other scattering processes $\Gamma^{(0\to 1)}=\Gamma^{(1\to 0)}=\Gamma^{(el)}=0$. In the presence of other scattering processes, there will be additional terms for other physical processes as we derive below. The contribution from leakage-only comes from subtracting out the ideal case where no qubits experience leakage, $\langle \sigma_j^+ \rangle_\text{leakage\ only} = (e^{-\lambda t}/2) \prod_{i \neq j}[I(J_{ij},t) + L(J_{ij},t)] - (e^{-\lambda t}/2)\prod_{i\neq j}I(J_{ij},t)$. 

The expectation value $\langle \sigma^-\rangle$ is computed similarly. Noting that this simply changes the sign of the $J_{ij}, J_{k_0j},$ and $J_{k_1j}$ in (\ref{sigma plus Heisenberg}) leads to the general formula for $\nu_j \in \{+,-\}$
\begin{align} \langle \sigma_j^{\nu_j} \rangle  =  \frac{e^{-\lambda t}}{2} \prod_{i\neq j} \left[I(\nu_jJ_{ij},t)  + L(\nu_jJ_{ij},t) \right] \end{align}
where the notation $\nu_j J_{ij}$ is used to denote signs of the $J_{ij}$. Expectation values of $m$ raising and lowering operators $\langle \prod_{j \in M} \sigma^{\nu_j}_j \rangle$ are also computed similarly. Noting that the Heisenberg representation analogous to (\ref{sigma plus Heisenberg}) now contains sums such as $\sum_{j\in M} \nu_j J_{ij}$, similar steps to above yield
\begin{align} \left\langle \prod_{j \in M} \sigma^{\nu_j}_j  \right\rangle  =  \frac{e^{-m\lambda t}}{2^m} \prod_{i \notin M} \left[I\left(\sum_{j \in M}\nu_jJ_{ij},t\right)  + L\left(\sum_{j \in M}\nu_jJ_{ij},t\right) \right] \end{align}

We now consider operator expectation values for projectors such as $\sigma^0_j = \ket{0_j}\bra{0_j}$.  Note this is an eigenstate of the $\sigma^z_j$ operators appearing in $\Hc(T_L)$, hence in the Heisenberg representation $\exp(i \Hc(T_L) t) \sigma^0_j \exp(-i \Hc(T_L) t) = \sigma^0_j.$  It follows that the expectation value $\bra{\overline \psi(t;T_L)}\sigma^0_j \ket{\overline \psi(t;T_L)}$ is separable. If qubit $j$ leaks to the ground state then  $\bra{\overline \psi(t;T_L)}\sigma^0_j \ket{\overline \psi(t;T_L)}=0$ while otherwise it evaluates to
\begin{equation} \bra{\overline \psi(t;T_L)}\sigma^0_j \ket{\overline \psi(t;T_L)} = \frac{e^{-\Gamma^{(0)}t}}{2} \langle \overline \psi_{i \neq j}(t;T_L)\ket{\overline \psi_{i \neq j}(t;T_L)}  \end{equation}
where $\ket{\overline \psi_{i \neq j}(t;T_L)}$ is the state on all qubits except $j$.  To compute $\langle \sigma^0_j\rangle$ we need to sum this over all trajectories $T_L$. From probability normalization it follows that $\sum_{T_L}  \langle \overline \psi_{i \neq j}(t;T_L)\ket{\overline \psi_{i \neq j}(t;T_L)} = 1$ and hence
\begin{equation} \langle \sigma^0_j \rangle = \sum_{T_L}  \bra{\overline \psi(t;T_L)}\sigma^0_j \ket{\overline \psi(t;T_L)}  = \frac{e^{-\Gamma^{(0)}t}}{2}\sum_{T_L}  \langle \overline \psi_{i \neq j}(t;T_L)\ket{\overline \psi_{i \neq j}(t;T_L)}  = I^{(0)}(0,t) \end{equation}
where $I^{(0)}(0,t) = e^{-\Gamma^{(0)}t}/2$ following Eq.~(\ref{I01}). Similar considerations give
\begin{equation} \langle \sigma^1_j \rangle = I^{(1)}(0,t) \end{equation}
and 
\begin{equation} \langle \sigma^g_j\rangle =L(0,t). \end{equation}

Finally we consider expectation values of products of different types of operators. Products of the projectors $\sigma^0_j$, $\sigma^1_j$, and $\sigma^g_j$ all commute with $\Hc(T_L)$ in the Heisenberg representation, so the expectation of their product is simply the product of their individual expectations.  Products of raising or lowering operators with projectors will contribute phases in the Heisenberg representation.  For example, $e^{i \Hc(T_L)t}  \sigma^0_k \sigma^+_j  e^{-i \Hc(T_L)t}$ yields a term similar to the right-side of (\ref{sigma plus Heisenberg}) except that $\sigma^z_k$ is replaced with its eigenvalue from $\sigma^0_k$. Similar considerations apply to other combinations of projectors and raising or lowering operators. For a generic operator
\begin{equation} \bm \sigma = \prod_{a \in P^{(0)}} \sigma_a^0 \prod_{b \in P^{(1)}} \sigma_b^1 \prod_{c \in P^{(g)}} \sigma_c^g \prod_{j \in M} \sigma_j^{\mu_j} \end{equation}
where $M, P^{(0)}, P^{(1)},$ and $P^{(g)}$ are sets of qubits, with $m$ is the number of qubits in the set $M$,
these considerations lead to the general operator expectation value formula
\begin{align} \label{generic expectation value leakage only} \left\langle \bm \sigma  \right\rangle = &\frac{e^{-m\lambda t}}{2^{m}} F^{(0)}(\bm \nu, \bm J,t) F^{(1)}(\bm \nu, \bm J,t) F^{(g)}(\bm \nu, \bm J,t)F(\bm \nu, \bm J,t). \end{align}
The functions $F$ are
\begin{align} F^{(0)}(\bm \nu, \bm J,t) & = \prod_{a \in P^{(0)}} \left[I^{(0)}\left(\sum_{j \in M}\nu_j J_{aj},t\right)\right] \nonumber\\
F^{(1)}(\bm \nu, \bm J,t) & = \prod_{b \in P^{(1)}} \left[I^{(1)}\left(\sum_{j \in M}\nu_j J_{bj},t\right)\right] \nonumber\\
F^{(g)}(\bm \nu, \bm J,t) & = \prod_{c \in P^{(g)}} \left[L\left(\sum_{j \in M}\nu_j J_{cj},t\right)\right] \nonumber\\
F(\bm \nu, \bm J,t) & = \prod_{i\notin M,P^{(0)},P^{(1)},P^{(g)}} \left[I\left(\sum_{j \in M}\nu_j J_{ij},t\right) + L\left(\sum_{j \in M}\nu_j J_{ij},t\right) \right]\end{align}
where $\bm \nu$ is a list of the $\nu_j \in \{+,-\}$ for qubits $j \in M$ and $\bm J$ is the set of coupling elements.

\subsection{Leakage with elastic scattering} 

We now consider elastic scattering processes $\Jc^{(el)}_j = \sqrt{\Gamma^{(el)}/8}\sigma^z_j$ in addition to leakage, following similar reasoning to Foss-Feig.~{\it et al.} \cite{Foss-Feig_PRA:2013,fossfeigdissertation}. The basic physical picture is that elastic scattering leads to dephasing when the scattered photons carry information about the state of the qubit (see Ref.~\cite{uys2010decoherence}).  The expectation $\langle \sigma^+_j\rangle$ depends on coherence between $\ket{0_j}$ and $\ket{1_j}$, hence it will be suppressed by the dephasing.  The expectation value also depends on the interaction with other qubits through the Ising interaction, but since this interaction depends only on populations in $\ket{0_i}$ and $\ket{1_i}$ of the other qubits, and not the phase coherence of these other qubits, we will show that elastic scattering on qubits $i\neq j$ does not influence the result for $\langle \sigma^+_j\rangle$.  This same result was derived in the context of Raman scattering in Ref.~\cite{Foss-Feig_PRA:2013,fossfeigdissertation}, where additional details can be found. 

A generic trajectory $T_{Le}$ containing both leakage and $\mu_i$ elastic scattering events on each qubit $i$ is 
\begin{align} \label{leakage elastic trajectory} \ket{\overline{\psi}(t;T_{Le})} = e^{-i \Hc(T_{Le})t} & \prod_{k_0 \in K_0} \frac{\sqrt{\Gamma^{(0\to g)} \delta t} e^{-t_{k_0} \Gamma^{(0)} /2}e^{-t_{k_0} \Gamma^{(el)} /8}(\sqrt{\Gamma^{(el)}\delta t}/2)^{\mu_{k_0}} \ket{g_{k_0}}}{\sqrt{2}} \nonumber\\
& \times \prod_{k_1 \in K_1} \frac{\sqrt{\Gamma^{(1\to g)} \delta t} e^{-t_{k_1} \Gamma^{(1)} /2}e^{-t_{k_1} \Gamma^{(el)} /8} (-\sqrt{\Gamma^{(el)}\delta t}/2)^{\mu_{k_1}}\ket{g_{k_1}}}{\sqrt{2}}
\nonumber\\
& \times \prod_{i \notin K} \left(\frac{\sqrt{\Gamma^{(el)}\delta t}\sigma^z_i}{2}\right)^{\mu_i}e^{-t \Gamma^{(el)} /8}\frac{e^{-\Gamma^{(0)}t/2}\ket{0_i} + e^{-\Gamma^{(1)}t/2}\ket{1_i}}{\sqrt{2}} \end{align}
and this gives
\begin{align} \label{generic  TLe} \bra{\overline{\psi}(t;T_{Le})}\sigma_j^+ \ket{\overline{\psi}(t;T_{Le})}   = & e^ {-\lambda t}\prod_{k_0 \in K_0} e^{-\Gamma^{(el)}t_{k_0}/4}\left(\frac{\Gamma^{(el)}\delta t}{4}\right)^{\mu_{k_0}} \frac{\Gamma^{(0\to g)} (\delta t)_{k_0} e^{it_{k_0} (2J_{k_0j}/N+i\Gamma^{(0)})}}{2} \nonumber\\
& \times \prod_{k_1 \in K_1} e^{-\Gamma^{(el)}t_{k_1}/4}\left(\frac{\Gamma^{(el)}\delta t}{4}\right)^{\mu_{k_1}} \frac{\Gamma^{(1\to g)} (\delta t)_{k_1} e^{-it_{k_1} (2J_{k_1j}/N-i\Gamma^{(1)}) }}{2} \nonumber\\
& \times \prod_{i \notin K} \frac{e^{-\Gamma^{(0)}t/2}\bra{0_i} + e^{-\Gamma^{(1)}t/2}\bra{0_i}}{\sqrt{2}} e^{-t \Gamma^{(el)} /8}\left(\frac{\sqrt{\Gamma^{(el)}\delta t}\sigma^z_i}{2}\right)^{\mu_i} \nonumber\\
& \times e^{i\Hc(T_{Le})t}\sigma_j^+e^{-i\Hc(T_{Le})t} \nonumber\\
& \times \prod_{i' \notin K}e^{-t \Gamma^{(el)} /8}\left(\frac{\sqrt{\Gamma^{(el)}\delta t}\sigma^z_{i'}}{2}\right)^{\mu_{i'}}  \frac{e^{-\Gamma^{(0)}t/2}\ket{0_{i'}} + e^{-\Gamma^{(1)}t/2}\ket{0_{i'}}}{\sqrt{2}}\end{align}
To compute $\langle \sigma^+_j\rangle$ we will need to sum $\bra{\overline{\psi}(t;T_{Le})}\sigma_j^+ \ket{\overline{\psi}(t;T_{Le})}$ over all possible sequences of elastic scattering events for a fixed set of leakage times and qubits.  We will then sum that result over all possible leakage trajectories, using methods developed in the previous section. 

First we will sum over all  possible times the elastic scattering could occur for given numbers of scattering events $\{\mu_i\}$.  This produces Riemann integrals which evaluate to 
\begin{equation} \int_0^tdt_{\mu} \int_0^{t_{\mu}} dt_{\mu-1} \ldots \int_0^{t_2} dt_1 = \frac{t^\mu}{\mu!}\end{equation}
Furthermore, we need to sum over all possible numbers of events $\mu_i$ for a given process.  For qubits $k \in K$ that leak from the qubit manfiold, and for qubits $i \neq j$ undergoing ideal evolution, this will produce factors such as 
\begin{equation} e^{-\Gamma^{(el)}t/4}\sum_{\mu=0}^\infty \frac{1}{\mu!}\left(\frac{\Gamma^{(el)}t}{4}\right)^{\mu} = 1 \end{equation} 
where the equality comes from recognizing that the sum is just the series expansion of $e^{\Gamma^{(el)}t/4}$. In the sum over all paths, this eliminates all elastic scattering contributions except ones on qubit $j$, where we will have a sum over terms
\begin{equation} \left(\frac{\sqrt{\Gamma^{(el)}\delta t}\sigma^z_j }{2}\right)^{\mu_j} \sigma_j^+ \left(\frac{\sqrt{\Gamma^{(el)}\delta t}\sigma^z_j}{2}\right)^{\mu_j} = \left(-\frac{\Gamma^{(el)}\delta t}{4}\right)^{\mu_j} \sigma^+_j \end{equation} 
Summing this over all times and all $\mu_j$ produces a factor
\begin{equation} e^{-\Gamma^{(el)}t/4}\sum_{\mu_j=0}^\infty \frac{1}{\mu_j!}\left(-\frac{\Gamma^{(el)}t}{4}\right)^{\mu_j} = e^{-\Gamma^{(el)}t/2} \end{equation} 
Hence the elastic scattering contributes a factor $e^{-\Gamma^{(el)}t/2}$ to $\langle \sigma^+_j\rangle$ that is associated with decoherence of qubit $j$, while for other qubits the elastic scattering does not affect the trajectory expectation value.  Hence, we can sum over all leakage trajectories as before to obtain
\begin{align} \label{sigma plus leakage elastic final} \langle \sigma_j^+ \rangle & = \sum_{T_{Le}}\bra{\overline{\psi}(t;T_{Le})}\sigma_j^+ \ket{\overline{\psi}(t;T_{Le})}  =  \frac{e^{-\Gamma t}}{2} \prod_{i\neq j} \left[I(J_{ij},t)  + L(J_{ij},t) \right] \end{align}
This is the same result as (\ref{sigma plus leakage final}) except with an additional factor $\Gamma^{(el)}/2$ in the leftmost exponent, with $\Gamma = \lambda + \Gamma^{(el)}/2$. Similar results hold for the other operator expectation values, as first described in Ref.~\cite{Foss-Feig_PRA:2013}. The analysis of this section and the previous section account for terms $L(J,t)$ arising from trajectories with leakage and elastic scattering in the general relation (\ref{generic expectation value}).

\subsection{Raman and elastic scattering} \label{Raman appendix}
 
The master equation with Raman and elastic scattering has been treated previously by Foss-Feig.~{\it et al.}~\cite{Foss-Feig_PRA:2013} who derived one- and two-spin operator expectation values; see also the related dissertation Ref.~\cite{fossfeigdissertation}. The Raman scattering trajectories present a substantial difficulty that was not present in the leakage errors presented in the last section.  The difficulty is that multiple, and potentially infinite, sequences of Raman transitions such as $\ket{0_i} \to \ket{1_i} \to \ket{0_i} \to \ldots$ are possible. Hence, the analytic calculation is more complicated, and we will only summarize the main points here, with further details in Refs.~\cite{Foss-Feig_PRA:2013,fossfeigdissertation}.  After summarizing the main points of their derivation, we will extend their results to compute $R^{(0)}(J,t)$ and $R^{(1)}(J,t)$ in (\ref{generic expectation value}).  The individual quantities $R^{(0)}(J,t)$ and $R^{(1)}(J,t)$ were not derived in Refs.~\cite{Foss-Feig_PRA:2013,fossfeigdissertation}, but these are essential in computing generic operator expectation values in (\ref{generic expectation value}), which have not been derived previously to our knowledge. 


In this subsection we focus on trajectories with Raman transitions only.  Analogous to (\ref{leakage trajectory}), we define an arbitrary Raman trajectory denoted by a parameter $T_R$ specifying  a set of Raman jump operators and producing a trajectory quantum state
\begin{align} \ket{\overline \psi(t;T_R)} = e^{-i \Hc(T_R) t} & \prod_{k_0 \in K_0} \sqrt{p_{k_0}^{(0)}(T_R)} \ket{0_{k_0}} \prod_{k_1 \in K_1} \sqrt{p_{k_1}^{(1)}(T_R)}  \ket{1_{k_1}} \prod_{i \notin K} \frac{e^{-\Gamma^{(0)}t/2}\ket{0_i} + e^{-\Gamma^{(1)}t/2}\ket{1_i}}{\sqrt{2}} \end{align}
where $K_0$ and $K_1$ are sets of qubits undergoing sequences of Raman transitions that end in the respective states $\ket{0}$ and $\ket{1}$ with $K= K_0 \cup K_1$. Refs.~\cite{Foss-Feig_PRA:2013,fossfeigdissertation} derived the probability densities $p_k^{(0)}(T_R)$ and $p_k^{(1)}(T_R)$ associated with the trajectory $T_R$.  In the case of an even number of scattering events 
\begin{align} p^{(0)}_k(T_R) = p_{k}^{(0 \to 0)}(\mu_k) & = \frac{e^{-(\lambda t + \Delta\tau_{k})}}{2}\left(\Gamma^{(0\to 1)}\Gamma^{(1\to 0)}(\delta t)^2\right)^{\mu_k+1} \nonumber\\
p^{(1)}_k(T_R) =p_{k}^{(1 \to 1)}(\mu_k) & = \frac{e^{-(\lambda t + \Delta\tau_{k})}}{2}\left(\Gamma^{(0\to 1)}\Gamma^{(1\to 0)}(\delta t)^2\right)^{\mu_k+1} \end{align}
while for an odd number of scattering events
\begin{align}
p^{(0)}_k(T_R) = p_{k}^{(1 \to 0)}(\mu_k) & = \frac{e^{-(\lambda t + \Delta\tau_{k})}}{2}\Gamma^{(1\to0)}\delta t\left(\Gamma^{(0\to 1)}\Gamma^{(1\to 0)}(\delta t)^2\right)^{\mu_k} \nonumber\\
p^{(1)}_k(T_R) = p_{k}^{(0 \to 1)}(\mu_k) & = \frac{e^{-(\lambda t + \Delta\tau_{k})}}{2}\Gamma^{(0\to1)}\delta t\left(\Gamma^{(0\to 1)}\Gamma^{(1\to 0)}(\delta t)^2\right)^{\mu_k}, \end{align}
where $\Delta = (\Gamma^{(0)}-\Gamma^{(1)})/2$ is the difference in scattering decoherence rates from $\ket{0}$ and $\ket{1}$, $\tau_k = t^{(0)}_k-t^{(1)}_k$ is the difference in time spent in $\ket{0}$ and $\ket{1}$ in the trajectory, and where we have parameterized the number of Raman transitions by $\mu_k = (N_k-1)/2$ for odd numbers of scattering events $N_k$ on qubit $k$ and $\mu_k=(N_k-2)/2$ for even numbers of scattering events.  Note that in relation to Refs.~\cite{Foss-Feig_PRA:2013,fossfeigdissertation}, our expressions for the decoherence rate $\lambda$ and decoherence rate difference $\Delta$ are based on total decays $\Gamma^{(0)}$ and $\Gamma^{(1)}$, which follows from our definition of $\Hc_\text{eff}$ in Eq.~(5) of our main paper, and generalizes the Raman-only decay rates used in their work. The effective Hamiltonian for the trajectory is 
\begin{equation} \label{HR leakage} \Hc(T_R) = \frac{1}{N}\sum_{i,j \notin K_0, K_1} J_{ij} \sigma^z_i \sigma^z_j + \frac{1}{N}\sum_{k \in K} (\tau_k/t) \sum_{i \notin K} J_{ik}   \sigma^z_i ,\end{equation}
where effective magnetic field terms depend on the time differences $\tau_k$.

We shall now compute $\bra{\overline \psi(t;T_R)} \sigma^+_j \ket{\overline \psi(t;T_R)}$ beginning with the Heisenberg representation of $\sigma^+_j$
\begin{equation} \label{Heisenberg sigma plus TR} e^{i \Hc(T_R)t} \sigma_j^+ e^{-i \Hc(T_R)t} = \sigma_j^+ e^{i2tN^{-1}\sum_{i \notin K} J_{ij} Z_i} e^{i2N^{-1}\sum_{k \in K} J_{jk}\tau_k} \end{equation}
Then 
\begin{equation} \label{sigma plus TR} \bra{\overline \psi(t;T_R)} \sigma^+_j \ket{\overline \psi(t;T_R)} = \frac{e^{-\lambda t}}{2}\prod_{k_0 \in K_0} p_{k_0}^{(0)}(T_R)e^{i2\tau_{k_0}J_{jk_0}/N} \prod_{k_1 \in K_1} p_{k_1}^{(1)}(T_R)e^{i2\tau_{k_1}J_{jk_1}/N} \prod_{i \notin K} I(J_{ij},t)\end{equation}
If we additionally include elastic scattering events, and sum over all numbers of elastic scattering events and times, we obtain the result 
\begin{equation} \label{sigma plus TRe} \bra{\overline \psi(t;T_{R}^{(el)})} \sigma^+_j \ket{\overline \psi(t;T_{R}^{(el)})} = \frac{e^{-\Gamma t}}{2}\prod_{k_0 \in K_0} p_{k_0}^{(0)}(T_R^{(el)})e^{i2\tau_{k_0}J_{jk_0}/N} \prod_{k_1 \in K_1} p_{k_1}^{(1)}(T_R^{(el)})e^{i2\tau_{k_1}J_{jk_1}/N} \prod_{i \notin K} I(J_{ij},t)\end{equation}
which includes an additional prefactor $\exp(-\Gamma^{(el)}t/2)$ relative to the previous expression, see the previous subsection and Refs.~\cite{Foss-Feig_PRA:2013,fossfeigdissertation} for details.

Note that so far we have not specified individual Raman scattering times, and have instead formulated everything in terms of the number of scattering events (through $\mu_k$) and the differences in time spent in $\ket{0}$ and $\ket{1}$ (through  $\tau_k$).  Although the result (\ref{sigma plus TR}) does not depend on the individual scattering times, we must nonetheless account for them when summing over all trajectories to obtain a final result for $\langle \sigma_j^+\rangle$.  Using a similar approach to Refs.~\cite{Foss-Feig_PRA:2013,fossfeigdissertation}, we shall do this by defining probability densities $P^{(a\to a)}(\mu,\tau)$ and $P^{(a \to b)}(\mu,\tau)$ in terms of the relevant variables $\mu$ and $\tau$, where the superscript $(a\to a)$ indicates that trajectories begin and end in the same state $\ket{a}$, while the superscript $(a\to b)$ indicates trajectories that begin and end in different states $\ket{a} \neq \ket{b}$.  Unlike Ref.~\cite{Foss-Feig_PRA:2013} we shall keep track of both the starting and ending states to distinguish different contributions that are necessary to give an account of all possible operator expectation values in (\ref{generic expectation value}).  

Summing over all possible scattering times consistent with a given $\mu$ and $\tau$, in the limit $\delta t\to 0$, in Ref.~\cite{Foss-Feig_PRA:2013} and Appendix H of \cite{fossfeigdissertation} Foss-Feig {\it et al.}~derived [with $\tilde p^{(a\to a)} = p^{(a\to a)}/(\delta t)^{N_k}$]
\begin{align} \int_0^t dt_{N_k} \ldots \int_0^{t_1} dt_{1}\tilde p^{(a\to a)} = \int_{-t}^t d\tau P^{(a\to a)}(\mu,\tau),
\end{align} 
with
\begin{align} P^{(a \to a)}(\mu,\tau) & = \frac{\Gamma^{(a \to b)}\Gamma^{(b \to a)}t^{(a)}}{4} \left(\frac{\Gamma^{(0\to 1)}\Gamma^{(1\to 0)}}{4}\right)^{\mu} e^{-(\lambda t + \Delta\tau)}\frac{(t^2-\tau^2)^\mu}{\mu!(\mu+1)!}.\end{align}
Similar expressions hold for cases with odd numbers of transitions
\begin{align}
P^{(a \to b)}(\mu,\tau) & = \frac{\Gamma^{(a \to b)}}{4} \left(\frac{\Gamma^{(0\to 1)}\Gamma^{(1\to 0)}}{4}\right)^{\mu} e^{-(\lambda t + \Delta\tau)}\frac{(t^2-\tau^2)^\mu}{(\mu!)^2}  \end{align}
To compute $\langle \sigma^+_j\rangle$ as the sum over all trajectories $T_R^{(el)}$ we need a final sum over all possible numbers of Raman transitions.  The result is 
\begin{align} \langle \sigma^+_j\rangle = \sum_{T_R^{(el)}}\bra{\overline \psi(t;T_R^{(el)})} \sigma^+_j \ket{\overline \psi(t;T_R^{(el)})} =  \frac{e^{-\Gamma t}}{2} \prod_{i \neq j} \left[R^{(0)}(J_{ij},t) + R^{(1)}(J_{ij},t) + I(J_{ij},t)\right] \end{align}
where
\begin{align} R^{(0)}(J_{ij},t) = \sum_{\mu=0}^\infty \int_{-t}^t d\tau (P^{(0 \to 0)} + P^{(1\to 0)})e^{i2\tau J_{ij}/N} \nonumber\\
R^{(1)}(J_{ij},t) = \sum_{\mu=0}^\infty \int_{-t}^t d\tau (P^{(1 \to 1)} + P^{(0\to 1)})e^{i2\tau J_{ij}/N} \end{align}
respectively describe contributions from all Raman scattering sequences that end in states $\ket{0}$ and $\ket{1}$. Evaluating these expressions is nontrivial and we present details in Appendix \ref{integral evaluation}.  The result is 
\begin{align} R^{(0)}(J_{ij},t) = \frac{e^{-\lambda t}}{2} \left[\cos(\zeta(J_{ij}) t) - \cos(s(J_{ij})t) + \Gamma^{(1\to0)}t \mathrm{sinc}(\zeta(J_{ij}) t) + is(J_{ij})t\left(\mathrm{sinc}(\zeta(J_{ij}) t) - \mathrm{sinc}(s(J_{ij})t)\right) \right], \nonumber\\
R^{(1)}(J_{ij},t) = \frac{e^{-\lambda t}}{2} \left[\cos(\zeta(J_{ij}) t) - \cos(s(J_{ij})t) + \Gamma^{(0\to1)}t \mathrm{sinc}(\zeta(J_{ij}) t) - is(J_{ij})t\left(\mathrm{sinc}(\zeta(J_{ij}) t) - \mathrm{sinc}(s(J_{ij})t)\right) \right]\end{align}
where $\zeta(J_{ij}) = \sqrt{s^2(J_{ij})-\Gamma^{(0\to 1)}\Gamma^{(1\to0)}}.$

To derive arbitrary operator expectation values we follow the same procedure as the previous section, by evaluating the Heisenberg representation of $\bm \sigma = \prod_{j \in M} \sigma_j^{\mu_j} \prod_{a \in P^{(0)}} \sigma_a^0 \prod_{b \in P^{(1)}} \sigma_b^1 \prod_{c \in P^{(g)}} \sigma_c^g $ in place of (\ref{Heisenberg sigma plus TR}).  All steps follow the same reasoning as in the leakage case. The final result is 
\begin{align} \label{generic expectation value TR} \left\langle \bm \sigma  \right\rangle = &\frac{e^{-m\Gamma t}}{2^m} F^{(0)}(\bm \nu, \bm J,t) F^{(1)}(\bm \nu, \bm J,t) F^{(g)}(\bm \nu, \bm J,t)F(\bm \nu, \bm J,t) \end{align}
with
\begin{align} F^{(0)}(\bm \nu, \bm J,t) & = \prod_{a \in P^{(0)}} \left[I^{(0)}\left(\sum_{j \in M}\nu_j J_{aj},t\right) + R^{(0)}\left(\sum_{j \in M}\nu_j J_{aj},t\right)\right] \nonumber\\
F^{(1)}(\bm \nu, \bm J,t) & = \prod_{b \in P^{(1)}} \left[I^{(1)}\left(\sum_{j \in M}\nu_j J_{bj},t\right) + R^{(1)}\left(\sum_{j \in M}\nu_j J_{bj},t\right)\right] \nonumber\\
F(\bm \nu, \bm J,t) & = \prod_{i\notin M,P^{(0)},P^{(1)},P^{(g)}} \left[I\left(\sum_{j \in M}\nu_j J_{ij},t\right) + R\left(\sum_{j \in M}\nu_j J_{ij},t\right) \right]\nonumber\\
F^{(g)}(\bm \nu, \bm J,t) & = 0 \end{align}
where $R(J,t) = R^{(0)}(J,t) + R^{(1)}(J,t).$ This result generalizes the one- and two-spin expectation values of Refs.~\cite{Foss-Feig_PRA:2013,fossfeigdissertation} to arbitrary numbers of spins, and accounts for terms $R^{(0)},R^{(1)}$ in (\ref{generic expectation value}).
 
\subsection{Trajectories with both Raman scattering and leakage}

Here we consider trajectories $T_B$ in which a sequence of Raman and elastic scattering transitions are followed by leakage outside the qubit manifold. The sum over all such trajectories must then include both sums over all possible sequences of Raman transitions as well as sums over all possible final leakage times.  These sums can be evaluated by integrating our previous Raman transition results $R^{(0)}, R^{(1)}$ over all possible leakage times to obtain terms
\begin{equation} B(J_{ij},t) = \Gamma^{(0 \to g)} \int_0^t dt' R^{(0)}(J_{ij},t') + \Gamma^{(1 \to g)} \int_0^t dt' R^{(1)}(J_{ij},t') \end{equation} 
To evaluate these integrals define
\begin{align} f(\Gamma,\chi,t)= \int_0^t dt' e^{(i\chi - \Gamma)t'} =  e^{ i(\chi + i\Gamma)t/2} t\mathrm{sinc}[(\chi + i\Gamma)t/2] \nonumber\\
a(\Gamma,\chi,t) = \int_0^t dt' e^{-\Gamma t'} \cos(\chi t) =  \frac{f(\Gamma,\chi,t) + f(\Gamma,-\chi,t)}{2} \nonumber\\
b(\Gamma,\chi,t) = \int_0^t dt' e^{-\Gamma t'} t\mathrm{sinc}(\chi t) =  -\frac{i}{\chi}\frac{f(\Gamma,\chi,t) - f(\Gamma,-\chi,t)}{2} \end{align}
The two terms we are integrating evaluate to
\begin{align} \int_0^t dt' R^{(0)}(J_{ij},t') = \frac{1}{2}\bigg(a(\lambda,\zeta(J_{ij}),t) - a(\lambda,s(J_{ij}),t) + \Gamma^{(1\to0)} b(\lambda,\zeta(J_{ij}),t) +is(J_{ij})[b(\lambda,\zeta(J_{ij}),t) - b(\lambda,s(J_{ij}),t)] \bigg)\nonumber\\
\int_0^t dt' R^{(1)}(J_{ij},t') = \frac{1}{2}\bigg(a(\lambda,\zeta(J_{ij}),t) - a(\lambda,s(J_{ij}),t) + \Gamma^{(0\to1)} b(\lambda,\zeta(J_{ij}),t) -is(J_{ij})[b(\lambda,\zeta(J_{ij}),t) - b(\lambda,s(J_{ij}),t)]\bigg) \end{align}
and summing we obtain the result
\begin{align} B(J_{ij},t) & = \Gamma^{(L)}[a(\lambda,\zeta,t) - a(\lambda,s,t)] + \Gamma^{(B)}b(\lambda,\zeta,t) + is\Delta^{(L)}[b(\lambda,\zeta,t)-b(\lambda,s,t)] \end{align}
where $\Gamma^{(B)}=(\Gamma^{(0\to 1)}\Gamma^{(1\to g)} + \Gamma^{(1\to 0)}\Gamma^{(0\to g)})/2$ and $\Delta^{(L)} = (\Gamma^{(0\to g)}-\Gamma^{(1\to g)})/2$.
Combining these with the previous expressions for trajectories with leakage or Raman scattering, and generalizing to products of many operators following the reasoning of Sec.~\ref{leakage appendix}, yields the final result (\ref{generic expectation value}). 

\subsection{Probability analysis for sequences of Raman transitions ending in $\ket{0}$ or $\ket{1}$} \label{integral evaluation}

Here we discuss how to evaluate the terms $R^{(0)}(J_{ij},t)$ and $R^{(1)}(J_{ij},t)$ from Sec.~\ref{Raman appendix}. Recall these are
\begin{align} R^{(0)}(J_{ij},t) = \sum_{\mu=0}^\infty \int_{-t}^t d\tau (P^{(0 \to 0)} + P^{(1\to 0)})e^{i2\tau J_{ij}/N} \nonumber\\
R^{(1)}(J_{ij},t) = \sum_{\mu=0}^\infty \int_{-t}^t d\tau (P^{(1 \to 1)} + P^{(0\to 1)})e^{i2\tau J_{ij}/N} \end{align}
with 
\begin{align} P^{(a \to a)}(\mu,\tau) & = \frac{\Gamma^{(a \to b)}\Gamma^{(b \to a)}t^{(a)}}{4} \left(\frac{\Gamma^{(0\to 1)}\Gamma^{(1\to 0)}}{4}\right)^{\mu} e^{-(\lambda t + \Delta\tau)}\frac{(t^2-\tau^2)^\mu}{\mu!(\mu+1)!} \nonumber\\
P^{(a \to b)}(\mu,\tau) & = \frac{\Gamma^{(a \to b)}}{4} \left(\frac{\Gamma^{(0\to 1)}\Gamma^{(1\to 0)}}{4}\right)^{\mu} e^{-(\lambda t + \Delta\tau)}\frac{(t^2-\tau^2)^\mu}{(\mu!)^2} \end{align}

Foss-Feig {\it et al.}~\cite{Foss-Feig_PRA:2013,fossfeigdissertation} evaluated terms similar to $R^{(0)}$ and $R^{(1)}$ above, except that they expressed these in terms of contributions from even numbers of transitions $P^{(0\to0)}+P^{(1\to 1)}$ and odd numbers of transitions $P^{(0\to1)}+P^{(1\to 0)}$, which was a useful distinction for their purposes.  For our treatment we must distinguish by the final state $\ket{0}$ or $\ket{1}$ to determine $R^{(0)}$ and $R^{(1)}$, respectively, to compute generic operator expectation values in (\ref{generic expectation value}).  To do this we will evaluate integrals over each $P$ separately. The integrals over $P^{(0\to 1)}$ and $P^{(1\to 0)}$ follow directly from the analysis of Refs.~\cite{Foss-Feig_PRA:2013,fossfeigdissertation} after noting that their contribution from odd terms comes from a sum of two components that are identical up to prefactors $\Gamma^{(0\to 1)}$ and $\Gamma^{(1\to 0)}.$ The result following from their work is
\begin{align} \sum_{\mu=0}^\infty \int_{-t}^t d\tau P^{(1\to 0)}(\mu,\tau)e^{i2\tau J_{ij}/N} = \frac{e^{-\lambda t}}{2}\Gamma^{(1\to 0)}t\mathrm{sinc}(\zeta t)\nonumber\\
\sum_{\mu=0}^\infty \int_{-t}^t d\tau P^{(0\to 1)}(\mu,\tau)e^{i2\tau J_{ij}/N} = \frac{e^{-\lambda t}}{2}\Gamma^{(0\to 1)}t\mathrm{sinc}(\zeta t)\label{odd integrals}\end{align}

We now evaluate the integrals over $P^{(0\to 0)}$ and $P^{(1\to 1)}$.  To begin we express these in terms of $t$ and $\tau$ as
\begin{align} 
P^{(0 \to 0)} = \frac{\Gamma^{(1 \to 0)}\Gamma^{(0 \to 1)}(t+\tau)}{8}  e^{-\lambda t} \frac{(\Gamma^{(0 \to 1)}\Gamma^{(1 \to 0)}/4)^\mu}{\mu!(\mu+1)!} e^{-\tau \Delta} (t^2-\tau^2)^\mu \nonumber\\
P^{(1 \to  1)} = \frac{\Gamma^{(1 \to 0)}\Gamma^{(0 \to 1)}(t-\tau)}{8}  e^{-\lambda t} \frac{(\Gamma^{(0 \to 1)}\Gamma^{(1 \to 0)}/4)^\mu}{\mu!(\mu+1)!} e^{-\tau \Delta} (t^2-\tau^2)^\mu \end{align}
where we used $t^{(0)}=(t+\tau)/2$ and $t^{(1)}=(t-\tau)/2$. We can separate the integration over these $P^{(a\to a)}$ into parts with $t$ in the leftmost fraction and parts with $\tau$ in the leftmost fraction.  Refs.~\cite{Foss-Feig_PRA:2013,fossfeigdissertation} 
 evaluated the parts with $t$ in the leftmost fraction and obtained the result
\begin{equation}
e^{-\lambda t} \frac{\Gamma^{(1 \to 0)}\Gamma^{(0 \to 1)}}{8}\sum_{\mu=0}^\infty \frac{(\Gamma^{(0 \to 1)}\Gamma^{(1 \to 0)}/4)^\mu}{\mu!(\mu+1)!} t\int_{-t}^t d\tau   e^{is\tau} (t^2-\tau^2)^\mu  = \frac{1}{2}e^{-\lambda t} (\cos(\zeta t) - \cos(st))
\end{equation}
where $\zeta = \sqrt{s^2-\Gamma^{(0\to 1)}\Gamma^{(1\to 0)}}$. 
 Here we must also evaluate 
\begin{equation} e^{-\lambda t} \frac{\Gamma^{(1 \to 0)}\Gamma^{(0 \to 1)}}{8}\sum_{\mu=0}^\infty \frac{(\Gamma^{(0 \to 1)}\Gamma^{(1 \to 0)}/4)^\mu}{\mu!(\mu+1)!} \int_{-t}^t d\tau   \tau e^{is\tau} (t^2-\tau^2)^\mu  \end{equation}

To evaluate the integral over $\tau$ we will need to peform some transformations. First note
\begin{equation} \frac{-1}{2(\mu+1)} \frac{d}{d\tau} (t^2-\tau^2)^{\mu+1} = \tau(t^2-\tau^2)^\mu \end{equation}
So we can simplify using integration by parts
\begin{align} \int_{-t}^t d\tau   \tau e^{is\tau} (t^2-\tau^2)^\mu & = \frac{-1}{2(\mu+1)} \int_{-t}^t d\tau    e^{is\tau} \frac{d}{d\tau} (t^2-\tau^2)^{\mu+1} \nonumber\\
& = \frac{-1}{2(\mu+1)}\left(\int_{-t}^td\tau \frac{d}{d\tau} \left[e^{is\tau } (t^2-\tau^2)^{\mu+1}\right] - is\int_{-t}^td\tau e^{is\tau} (t^2-\tau^2)^{\mu+1} \right) \nonumber\\
& =  \frac{is}{2(\mu+1)}\int_{-t}^td\tau e^{is\tau} (t^2-\tau^2)^{\mu+1} \nonumber\\
& = ist(2t)^{\mu+1} \frac{j_{\mu+1}(st)\mu!}{s^{\mu+1}}\end{align}
where the final equality uses
\begin{equation} \int_{-t}^t d\tau (t^2-\tau^2)^{\mu+1} e^{is\tau} = (2t)^{\mu+2} \frac{j_{\mu+1}(st)(\mu+1)!}{s^{\mu+1}} \end{equation} 
from (B6) in Ref.~\cite{Foss-Feig_PRA:2013}, where $j_{\mu+1}$ is the $(\mu+1)$-order spherical Bessel function. Putting this into the full term 
\begin{equation} e^{-\lambda t} \frac{\Gamma^{(1 \to 0)}\Gamma^{(0 \to 1)}}{8}\sum_{\mu=0}^\infty \frac{(\Gamma^{(0 \to 1)}\Gamma^{(1 \to 0)}/4)^\mu}{\mu!(\mu+1)!} \int_{-t}^t d\tau   \tau e^{is\tau} (t^2-\tau^2)^\mu = \frac{ist}{2}e^{-\lambda t} \sum_{\mu=0}^\infty \frac{(\Gamma^{(0 \to 1)}\Gamma^{(1 \to 0)}t/2s)^{\mu+1}}{(\mu+1)!} j_{\mu+1}(st) \end{equation}
To evaluate this sum use $r = \Gamma^{(0 \to 1)}\Gamma^{(1 \to 0)}$ and
\begin{equation} \frac{2s}{t}\frac{d}{dr} \frac{(rt/2s)^{\mu+2}}{(\mu+2)!} = \frac{(rt/2s)^{\mu+1}}{(\mu+1)!} \end{equation}
giving
\begin{align} \frac{ist}{2}e^{-\lambda t} \sum_{\mu=0}^\infty \frac{(rt/2s)^{\mu+1}}{(\mu+1)!} j_{\mu+1}(st) & = \frac{d}{dr} is^2e^{-\lambda t} \sum_{\mu=0}^\infty \frac{(rt/2s)^{\mu+2}}{(\mu+2)!} j_{\mu+1}(st) \nonumber\\
& = \frac{d}{dr} is^2e^{-\lambda t} \left( -j_{-1}(st) - \frac{rt}{2s}j_0(st) + \sum_{\mu=0}^\infty \frac{(rt/2s)^{\mu}}{\mu!} j_{\mu-1}(st)\right) \nonumber\\
& = is^2e^{-\lambda t} \left(\frac{-t}{2s}j_0(st) + \frac{d}{dr} \frac{1}{st}\cos(\zeta t)\right)\nonumber\\
& = is^2e^{-\lambda t} \left(\frac{-t}{2s}j_0(st) + \frac{t}{2s}\sinc(\zeta t)\right)\nonumber\\
& = \frac{ist}{2}e^{-\lambda t} \left(\sinc(\zeta t) - \sinc(st)\right)\end{align} 
where we used the generating functions of the spherical Bessel functions $\sum_{\mu=0}^\infty \frac{(rt/2s)^{\mu}}{\mu!} j_{\mu-1}(st) = \cos(\zeta t)/st$. In total, this gives 
\begin{equation} \sum_{\mu=0}^\infty \int_{-t}^t d\tau P^{(0\to 0)}(\mu,\tau)e^{i2J_{ij}\tau} = \frac{e^{-\lambda t}}{2} \left[\cos(\zeta t) - \cos(st) + ist\left(\mathrm{sinc}(\zeta t) - \mathrm{sinc}(st)\right) \right] \end{equation}
\begin{equation} \sum_{\mu=0}^\infty \int_{-t}^t d\tau P^{(1\to 1)}(\mu,\tau)e^{i2J_{ij}\tau} = \frac{e^{-\lambda t}}{2} \left[\cos(\zeta t) - \cos(st) - ist\left(\mathrm{sinc}(\zeta t) - \mathrm{sinc}(st)\right) \right] \end{equation}
Combining these with the previous evaluations over odd numbers of transitions (\ref{odd integrals}) we obtain $R^{(0)}(J_{ij},t)$ and $R^{(1)}(J_{ij},t)$ in (\ref{R01}). 

\section{$^{40}$Ca$^+$ modelling}

Here we consider the specific setup with $^{40}$Ca$^+$ in the Penning trap operated at GTRI, see the main paper for details. Our analysis up until now has considered Ising dynamics in the presence of light scattering.  Similar models of Ising-dynamics with light scattering were used to successfully describe experiments such as Ref.~\cite{Bohnet_Science:2016}.  However, in a real experiment, the effective Ising evolution is generated using a spin-echo sequence, and modelling the light scattering dynamics including this spin-echo sequence has not yet been done to our knowledge.  To improve the physical realism of our modelling, here we treat the full spin-echo sequence generating Ising evolution for the proposed $^{40}$Ca$^+$ experiments.  We also summarize results of 2D ion crystal simulations that determine the Ising couplings $J_{ij}$.

\subsection{Ideal Dynamics}

The effective Hamiltonian for the chosen experimental setup (Fig.~1(a) in the main paper) is
\begin{equation} \label{H Ca} \Hc_\text{arm} = \frac{1}{N}\sum_{i<j} J_{ij} \ket{0_i 0_j}\bra{0_i 0_j} \end{equation}
where
\begin{equation} J_{ij} = N\Omega_i \Omega_j\sum_{m=0}^{N-1} \frac{\eta_{i,m}\eta_{j,m}\omega_m}{\mu^2-\omega_m^2}\end{equation}%

Evolving a state under $\Hc_\text{arm}$, in sequence with a spin-echo sequence with two $\pi$ pulses $S = \prod_i (\sigma^x_i + \ket{g_i}\bra{g_i})$, gives a total unitary
\begin{equation} \label{U spin echo} U = Se^{-i \Hc_\text{arm} t_\text{arm}}Se^{-i \Hc_\text{arm} t_\text{arm}} = e^{-i(S\Hc_\text{arm} S + \Hc_\text{arm})t_\text{arm}} = e^{-i \Hc_\text{se} t_\text{arm}} \end{equation}
where $t_\text{arm}$ is the time in each arm of the spin echo.  The ideal Hamiltonian evolution is then
\begin{align} \Hc_\mathrm{se} = S\Hc_\text{arm} S + \Hc_\text{arm} = \frac{1}{N}\sum_{i<j} J_{ij} (\ket{0_i 0_j}\bra{0_i 0_j} + \ket{1_i 1_j}\bra{1_i 1_j}) \end{align}
The operators on the right can be written in terms of Pauli-$Z$ operators with 
\begin{equation} \ket{0_i 0_j}\bra{0_i 0_j} + \ket{1_i 1_j}\bra{1_i 1_j} = \frac{1}{2}\left(\sigma^z_i \sigma^z_j + \mathbb{1}_{ij}\right)\end{equation}
where $\mathbb{1}_{ij}$ is the identity within the qubit manifold of ions $i$ and $j$. The term $\mathbb{1}_{ij}$ will contribute a global phase across the qubit states, which will not effect their dynamics. Hence, we take the  spin-echo Hamiltonian as
\begin{equation} \Hc_\text{se} = \frac{1}{2N} \sum_{i<j} J_{ij} \sigma^z_i \sigma^z_j \end{equation}
Taking $J_{ij}$ as the Ising couplings, this reduces to 
\begin{equation} \label{Ca H Hising} \Hc_\text{se} = \frac{1}{2} \Hc_\text{Ising} \end{equation} 
where $\Hc_\text{Ising} = \Hc = N^{-1}\sum_{i < j}J_{ij} \sigma_i^z \sigma^z_j$ is the Ising Hamiltonian in Eq.~(2) of the main text. Hence if we evolve in each arm of the spin echo sequence for time $t_\text{arm} = 2t$, then we will produce the desired Ising evolution $\exp(-i \Hc t)$.  The spin-echo includes two arms, so the actual experimental runtime is $t_\text{expt} = 2t_\text{arm} = 4t$ \cite{mcmahon_individual_2024}.

\begin{figure}    
    \includegraphics[height=6cm,width=\columnwidth,keepaspectratio]{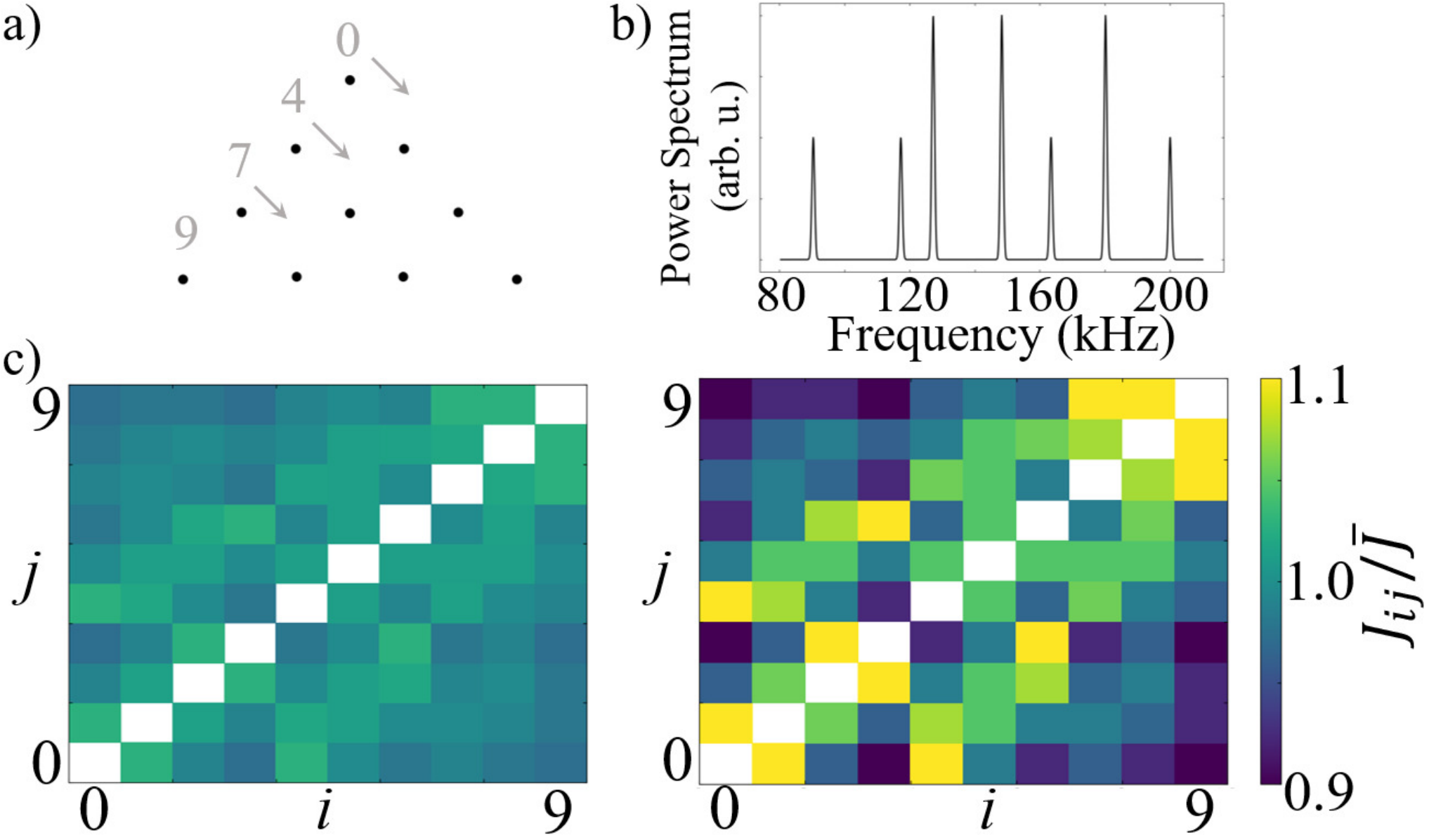}
    \caption{(a) Simulation of 2D ion array, which is a triangular lattice in the rotating frame as in Ref.~\cite{mcmahon_individual_2024}. (b) Simulated power spectrum of collective axial eigenmodes of motion for the crystal in (a). (c) Simulated $J_{ij}/\bar{J}$ spin couplings for the crystal in (a) assuming a COM mode detuning of $\delta = 2\pi\times 0.5$~kHz (left) and $\delta = 2\pi\times 2$~kHz (right). }
    \label{fig:crystal simulation}
\end{figure}

Figure \ref{fig:crystal simulation} shows details of a 2D ion array simulation, including the triangular crystal lattice, motional eigenmodes, and couplings $J_{ij}/N$.

\subsection{Light scattering master equation for $^{40}$Ca$^+$}

Here we extend the previous analysis to model light scattering dynamics in the spin-echo sequence of the proposed experiments with $^{40}$Ca$^+$.  This includes the Hamiltonian (\ref{H Ca}) and the spin-echo sequence (\ref{U spin echo}). The effective Hamiltonian including dissipation is $\Hc_\text{eff} = N^{-1}\sum_{i<j} J_{ij} \ket{0_i0_j}\bra{0_i0_j}-i\sum_{\alpha,j} {\Jc^{\alpha}_j}^\dag\Jc^{\alpha}_j$. Including the spin echo operators gives
\begin{equation} S e^{-i \Hc_\text{eff}t_\text{arm}} S e^{-i \Hc_\text{eff}t_\text{arm}} = e^{-i t_\text{arm}(2N)^{-1} \sum_{i<j} J_{ij} \sigma_i^z \sigma_j^z} e^{-\Gamma^{(0)}t_\text{arm}(\ket{0_i}\bra{0_i}+\ket{1_i}\bra{1_i})} \end{equation}
noting that $\Gamma^{(1)}=0$ due to selection rules, see Fig.~1(a) in the main paper. For this setup there are seven possible types of trajectories depicted in Fig.~\ref{Ca scatter diagram}; the number of trajectories is finite because $\Gamma^{(1)}=0$ does not allow multiple scattering events in a single arm of the spin-echo.

\begin{figure}
    \includegraphics[height=12cm,width=14cm,keepaspectratio]{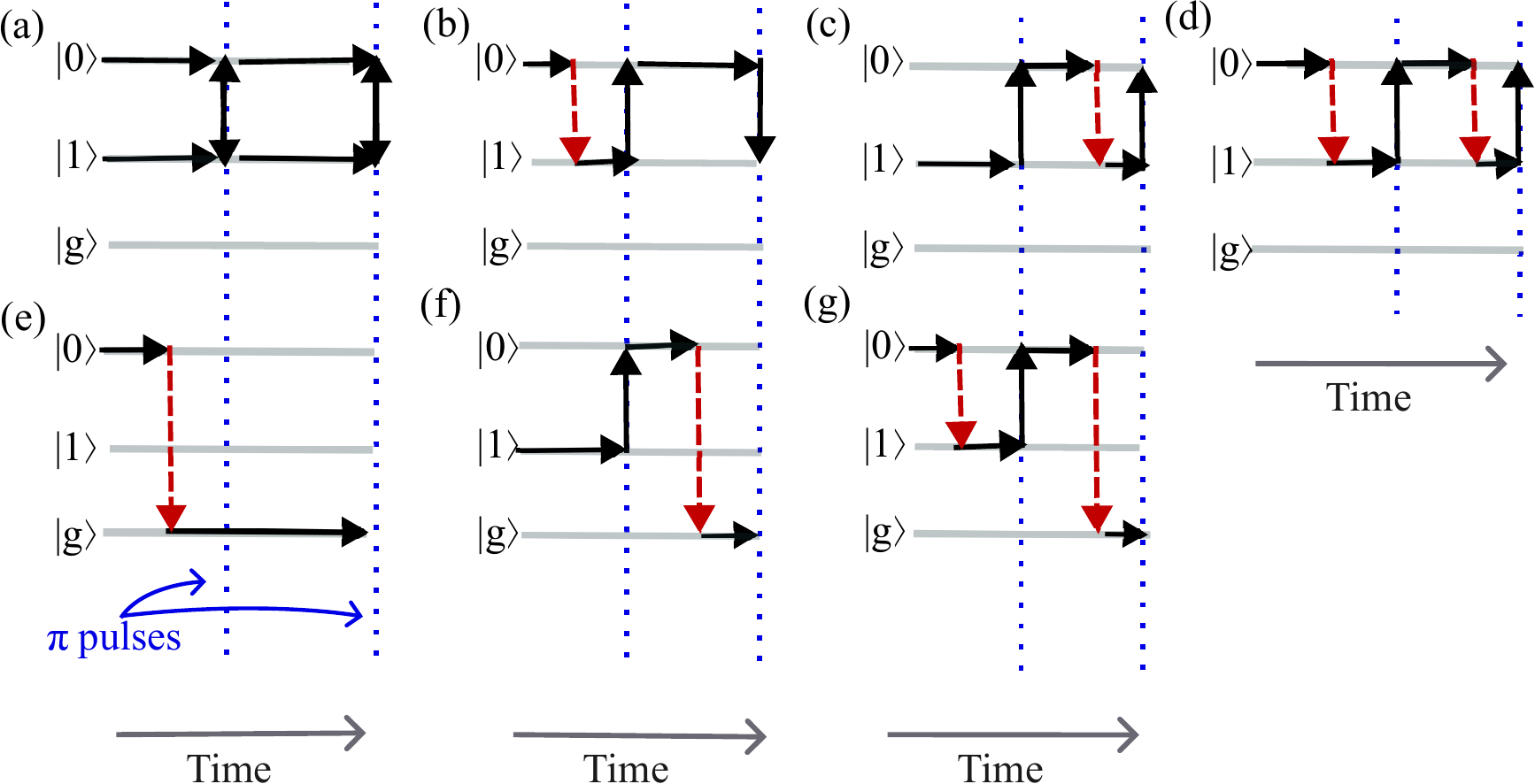}
    \caption{All possible types of trajectories in the model of $^{40}$Ca$^+$.  Black solid arrows show dynamics of populations in $\ket{0},\ket{1},$ and $\ket{g}$, red dashed arrows show light scattering transitions, and blue dotted lines show $\pi$-pulses $S$ used in the spin-echo. The seven possible types of dynamics are (a) the ideal trajectory, (b-d) Raman scattering trajectories, (e-f) leakage trajectories, and (g) a trajectory with both Raman scattering and leakage. }
    \label{Ca scatter diagram}
\end{figure}

We derived operator expectation values for this setup, yielding
\begin{align} \label{generic expectation value Ca} \left\langle \prod_{a \in P^{(0)}} \sigma_a^0 \prod_{b \in P^{(1)}} \sigma_b^1 \prod_{c \in P^{(g)}} \sigma_c^g \prod_{j \in M} \sigma_j^{\mu_j} \right\rangle = &\frac{e^{-m\tilde \Gamma t_\text{arm}}}{2^m} \tilde F^{(0)}(\bm \nu, \bm J,t_\text{arm}) \tilde F^{(1)}(\bm \nu, \bm J,t_\text{arm}) \tilde F^{(g)}(\bm \nu, \bm J,t_\text{arm})\tilde F(\bm \nu, \bm J,t_\text{arm}) \end{align}
where $\tilde \Gamma = \Gamma^{(0)} + \Gamma^{(el)}$ and 
\begin{align} \tilde F^{(0)}(\bm \nu, \bm J,t_\text{arm}) & = \prod_{a \in P^{(0)}} \left[\tilde I^{(0)}\left(J_a^{(M,\bm \nu)},t_\text{arm}\right) + \tilde R^{(0)}\left(J_a^{(M,\bm \nu)},t_\text{arm}\right)\right] \nonumber\\
\tilde F^{(1)}(\bm \nu, \bm J,t_\text{arm}) & = \prod_{b \in P^{(1)}} \left[\tilde I^{(1)}\left(J_b^{(M,\bm \nu)},t_\text{arm}\right) + \tilde R^{(1)}\left(J_b^{(M,\bm \nu)},t_\text{arm}\right)\right] \nonumber\\
\tilde F^{(g)}(\bm \nu, \bm J,t_\text{arm}) & = \prod_{c \in P^{(g)}} \left[\tilde L\left(J_c^{(M,\bm \nu)},t_\text{arm}\right) + \tilde B\left(J_c^{(M,\bm \nu)},t_\text{arm}\right)\right] \nonumber\\
\tilde F(\bm \nu, \bm J,t_\text{arm}) & = \prod_{i\notin M,P^{(0)},P^{(1)},P^{(g)}} \left[\tilde I\left(J_i^{(M,\bm \nu)},t_\text{arm}\right) + \tilde R\left(J_i^{(M,\bm \nu)},t_\text{arm}\right) + \tilde L\left(J_i^{(M,\bm \nu)},t_\text{arm}\right) + \tilde B\left(J_i^{(M,\bm \nu)},t_\text{arm}\right)\right].\end{align}
where $J_i^{(M,\bm \nu)} = \sum_{j \in M} \nu_j J_{ij}$. This is analogous to (\ref{generic expectation value}) but with the following functions denoted with a $\sim$ on the top

\begin{align} \tilde I^{(0)}(J,t_\text{arm}) = \frac{e^{iJt_\text{arm}/N} e^{-\Gamma^{(0)}t_\text{arm}}}{2}, \nonumber\\
\tilde I^{(1)}(J,t_\text{arm}) = \frac{e^{-iJt_\text{arm}/N} e^{-\Gamma^{(0)}t_\text{arm}}}{2} \nonumber\\
\tilde R^{(0)}(J,t_\text{arm}) = \frac{\Gamma^{(0\to 1)}f(\Gamma^{(0)},-J/N,t_\text{arm}) + (\Gamma^{(0\to 1)})^2f(\Gamma^{(0)},J/N,t_\text{arm})f(\Gamma^{(0)},-J/N,t_\text{arm})}{2}\nonumber\\
\tilde R^{(1)}(J,t_\text{arm}) = \frac{ \Gamma^{(0\to 1)}f(\Gamma^{(0)},J/N,t_\text{arm})e^{-(iJ/N+\Gamma^{(0)})t_\text{arm}}}{2}, \nonumber\\
\tilde L(J,t_\text{arm}) = \frac{\Gamma^{(0\to g)}[f(\Gamma^{(0)},J/N,t_\text{arm})+ f(\Gamma^{(0)},-J/N,t_\text{arm})]}{2}\nonumber\\
\tilde B(J,t_\text{arm}) = \frac{\Gamma^{(0\to 1)}\Gamma^{(0\to g)} f(\Gamma^{(0)},-J/N,t_\text{arm}) f(\Gamma^{(0)},J/N,t_\text{arm})}{2}\end{align}
The expression for $\tilde L$ is similar to the one described in Sec.~\ref{results summary}, except with a factor $1/2$ difference on $J$ due to the 1/2 factor in (\ref{Ca H Hising}), while expressions for $\tilde R$ and $\tilde B$ differ more significantly from the $R$ and $B$ in Sec.~\ref{results summary} due in part to the relatively limited sets of Raman transitions that are possible in Fig.~\ref{Ca scatter diagram}. We present derivations for the ideal and leakage trajectories below, expressions for the $\tilde R$ and $\tilde B$ follow from very similar considerations.

\subsubsection{Ideal trajectory}
The ideal trajectory $\ket{\psi(t;T_I)}$ evolves under
\begin{equation} U = S e^{-i \Hc_\text{eff} t_\text{arm}} S e^{-i \Hc_\text{eff} t_\text{arm}} \end{equation} 
 giving
\begin{equation} \ket{\overline\psi(T_I)} = e^{-i \Hc_\text{se}t_\text{arm}} \prod_i e^{-\Gamma^{(0)}t_\text{arm}/2}\frac{\ket{0_i} + \ket{1_i}}{\sqrt{2}} \end{equation}
The Heisenberg representation for $\sigma^+_i = \ket{0_i}\bra{1_i}$ gives 
\begin{equation} e^{i\Hc_\text{se}t_\text{arm}} \ket{0_i}\bra{1_i} e^{-i\Hc_\text{se}t_\text{arm}} = e^{it_\text{arm}N^{-1}\sum_{j \neq i} J_{ij} \sigma^z_j} \ket{0_i}\bra{1_i} \end{equation}
and the expectation value for the ideal trajectory is
\begin{equation} \langle\overline \psi(t;T_I)\vert \sigma^+_i \ket{\overline\psi(t;T_I)} = e^{-\Gamma^{(0)}t_\text{arm}} \prod_{j \neq i} (\tilde I^{(0)}(J_{ij},t_\text{arm}) + \tilde I^{(1)}(J_{ij},t_\text{arm}))\end{equation}
where
\begin{equation} \tilde I^{(0)}(J_{ij},t_\text{arm}) = \frac{e^{-\Gamma^{(0)}t_\text{arm}}e^{iJ_{ij}t_\text{arm}/N}}{2}, \ \ \ \tilde I^{(1)}(J_{ij},t) = \frac{e^{-\Gamma^{(0)}t_\text{arm}}e^{-iJ_{ij}t_\text{arm}/N}}{2}, \end{equation}
\begin{equation}  \tilde I(J,t_\text{arm}) = \tilde I^{(0)}(J,t_\text{arm})+\tilde I^{(1)}(J,t_\text{arm})=e^{-\Gamma^{(0)}t_\text{arm}}\cos(J_{ij}t_\text{arm}/N)\end{equation}

\subsubsection{$\ket{0_\alpha} \to \ket{g_\alpha}$ trajectory}

A trajectory with decay $\ket{0_\alpha} \to \ket{g_\alpha}$ will evolve under 
\begin{equation}  S e^{-i \Hc_\text{se} t_\text{arm}} S e^{-i \Hc_\text{se}(t_\text{arm}-t')} \sqrt{\Gamma^{(0\to g)} \delta t} \ket{g_\alpha} \bra{0_\alpha} e^{-i \Hc_\text{se} t'}  \end{equation}
Similar to the previous analysis, we can take the qubit $\alpha$ as beginning in $\ket{g_\alpha}$ and evolve instead with 
\begin{equation} U = e^{-it_\text{arm}( S\tilde \Hc_\text{se}(T)S + \tilde \Hc_\text{se}(T))} e^{-it' N^{-1}\sum_{j \neq \alpha} J_{\alpha,j} \ket{0_j}\bra{0_j} }\sqrt{\Gamma^{(0\to g)} \delta t} \end{equation}
where $\tilde \Hc_\text{se}$ ignores the contributions from qubit $\alpha$, which are accounted for by the effective field $e^{-it' \sum_{j \neq \alpha} J_{\alpha,j} \ket{0_j}\bra{0_j} }$. The Heisenberg representation for $\sigma_i^+$ gives
\begin{equation} U^\dag \sigma_i^+ U = \Gamma^{(0\to g)} \delta t \ket{0_i}\bra{1_i} e^{it'J_{\alpha i}/N} e^{it_\text{arm}N^{-1}\sum_{j \neq i,\alpha} J_{i,j}\sigma^z_j} \end{equation}
while for the effective initial state we have 
\begin{equation} \ket{\overline\psi(0;T)} = \frac{e^{-\Gamma^{(0)}t'/2}}{\sqrt{2}}\ket{g_\alpha} \prod_{j \neq \alpha} e^{-\Gamma^{(0)}t_\text{arm}/2} \frac{\ket{0_j} + \ket{1_j}}{\sqrt{2}}\end{equation}
The expectation of $\sigma_i^+$ is
\begin{equation} \bra{\overline\psi(0;T)}U^\dag \sigma^+_i U \ket{\overline\psi(0;T)} = \frac{\Gamma^{(0\to g)}\delta te^{(iJ_{\alpha i}/N - \Gamma^{(0)})t'}}{2} \frac{e^{-\Gamma^{(0)}t_\text{arm}}}{2} \prod_{j \neq \alpha,i} I(J_{ij},t_\text{arm})  \end{equation}
To obtain the contribution from all such trajectories we need to integrate over all $t'$
\begin{equation} \int_0^tdt' \bra{\overline\psi(0;T)}U^\dag \sigma^+_i U \ket{\overline\psi(0;T)}  = \tilde L^{(0\to g)}(J_{\alpha i},t_\text{arm}) \frac{e^{-\Gamma^{(0)}t_\text{arm}}}{2} \prod_{j \neq \alpha,i} I(J_{ij},t_\text{arm})   \end{equation}
where
\begin{equation} \tilde L^{(0\to g)}(J_{\alpha i},t_\text{arm}) = \frac{\Gamma^{(0\to g)}f(\Gamma^{(0)},J_{\alpha i}/N,t_\text{arm})}{2} \end{equation}

\subsubsection{$\ket{1_\alpha} \to \ket{g_\alpha}$}

Following similar steps to before, we begin with the effective initial state 
\begin{equation} \ket{\psi(0)} = e^{-\Gamma^{(0)}t'/2} \ket{g_\alpha} \prod_{j \neq \alpha} e^{-\Gamma^{(0)}t_\text{arm}/2}\frac{\ket{0_j}+\ket{1_j}}{\sqrt{2}}, \end{equation}
the effective unitary dynamics
\begin{equation} U = \sqrt{\Gamma^{(0\to g)}\delta t}e^{-i (S\tilde \Hc S + \tilde \Hc)t} e^{-it'N^{-1}\sum_{j \neq \alpha} J_{\alpha j} \ket{1_j}\bra{1_j}}, \end{equation}
the Heisenberg $\sigma_i^+$ 
\begin{equation} U^\dag \sigma^+_iU = \ket{0_i}\bra{1_i} e^{-it'J_{\alpha i}/N}e^{i t_\text{arm}N^{-1}\sum_{j \neq \alpha,i} J_{ij} \sigma^z_j}, \end{equation} 
the expectation value for a single trajectory
\begin{equation} \bra{\overline\psi(0;T)}U^\dag \sigma^+_i U \ket{\overline\psi(0;T)} = \frac{\Gamma^{(0\to g)}\delta t e^{-t'(iJ_{\alpha,i}/N+\Gamma^{(0)})}}{2}\frac{e^{-\Gamma^{(0)}t_\text{arm}}}{2}\prod_{j \neq \alpha} I(J_{ij},t_\text{arm}), \end{equation}
and then integrating over all trajectories
\begin{equation} \int_0^t dt' \bra{\overline\psi(0;T)}U^\dag \sigma^+_i U \ket{\overline\psi(0;T)} =  L^{(1 \to g)}(J_{\alpha i},t) \frac{e^{-\Gamma^{(0)}t_\text{arm}}}{2}\prod_{j \neq \alpha}  I(J_{ij},t_\text{arm}), \end{equation}
with
\begin{equation} \tilde L^{(1 \to g)}(J_{\alpha i},t_\text{arm}) = \frac{\Gamma^{(0\to g)}f(\Gamma^{(0)},-J_{\alpha i},t_\text{arm})}{2} \end{equation}
The total leakage term is the sum of the two previous terms 
\begin{equation} \tilde L(J_{\alpha i},t_\text{arm}) = \tilde L^{(0\to g)}(J_{\alpha i},t_\text{arm}) + \tilde L^{(1\to g)}(J_{\alpha i},t_\text{arm}) \end{equation} 
Other terms are derived similarly.

\section{GHZ state fidelity model for unequal $J_{ij}$}

To understand how unequal $J_{ij}$ affects the GHZ state preparation fidelity we consider noiseless evolution under an Ising Hamiltonian
\begin{equation} \label{HIsing supplemental} \Hc = \frac{1}{N}\sum_{i < j} J_{ij} \sigma^z_i\sigma^z_j\end{equation}
and express this in terms of the average coupling $J$ and the $(i,j)-$dependent deviations from equal coupling $\delta_{ij}$, 
\begin{equation} \Hc =  \frac{1}{N}\sum_{i < j} (J+\delta_{ij}) \sigma^z_i\sigma^z_j\end{equation}

Now we want to compare states with an equal coupling $\Hc_\mathrm{equal} = J\sum_{i < j} \sigma^z_i\sigma^z_j$ to the states that are produced with $\Hc$.  These states are
\begin{equation} \ket{\psi_\mathrm{equal}} = e^{-i Jt N^{-1}\sum_{i<j} \sigma^z_i\sigma^z_j} \ket{\psi_0} \end{equation}
and
\begin{equation} \ket{\psi_\mathrm{unequal}} = e^{-i t N^{-1}\sum_{i<j}(J + \delta_{ij}) \sigma^z_i\sigma^z_j} \ket{\psi_0} \end{equation}
respectively, where
\begin{equation} \ket{\psi_0} = \ket{+}^{\otimes N}.\end{equation}
The error from the unequal coupling can be quantified by the fidelity
\begin{equation} F_\text{unequal} = |\langle \psi_\mathrm{equal}\ket{\psi_\mathrm{unequal}}|^2 = |\bra{\psi_0} e^{i tN^{-1}\sum_{i < j} \delta_{ij}\sigma^z_i\sigma^z_j} \ket{\psi_0}|^2\end{equation}
We give two different derivations of $F_\text{unequal}$ in the next two subsections.  The first is simple and intuitive, the second provides systematic extensions to higher-order corrections. 

\subsection{First derivation}

We will need to evaluate
\begin{equation} \bra{\psi_0} e^{i tN^{-1}\sum_{i < j} \delta_{ij}\sigma^z_i\sigma^z_j} \ket{\psi_0} = \frac{1}{2^n}\sum_{z=0}^{2^n-1} \langle z \vert e^{i tN^{-1}\sum_{i < j} \delta_{ij}\sigma^z_i\sigma^z_j} \ket{z}  = \frac{1}{2^n}\sum_{z=0}^{2^n-1}  e^{i tN^{-1}\sum_{i < j} \delta_{ij}\zeta_i(z)\zeta_j(z)}  \end{equation}
where $\zeta_i(z),\zeta_j(z) = \pm 1$ are the respective eigenvalues of $\sigma^z_i, \sigma^z_j$ for the state $\ket{z}$. This is an average of $e^{i t\sum_{i < j} \delta_{ij}\zeta_i(z)\zeta_j(z)}$ over all $\ket{z}$. For $z$ selected uniformly at random, the phase
\begin{equation} \phi(z) = \frac{t}{N}\sum_{i < j} \delta_{ij}\zeta_i(z)\zeta_j(z) \end{equation}
is a sum of random variables, and so we expect $\phi(z)$ should behave as a random Gaussian variate following the central limit theorem. The mean of $\phi(z)$ is zero since $\zeta_i(z)\zeta_j(z) = \pm 1$ with equal probability, while the variance is
\begin{equation} \sigma^2 = \frac{t^2}{N^2}\sum_{i < j} \delta_{ij}^2 \end{equation}
The sum can then be approximated as a Gaussian integral
\begin{equation} \frac{1}{2^n}\sum_{z=0}^{2^n-1}  e^{i tN^{-1}\sum_{i < j} \delta_{ij}\zeta_i(z)\zeta_j(z)} \approx \frac{1}{\sqrt{2\pi}\sigma} \int_{-\infty}^\infty d\phi e^{-\phi^2/2\sigma^2} e^{i\phi} = e^{-\sigma^2/2} \end{equation}
Squaring this gives the approximate fidelity
\begin{equation} \label{F approx} F_\text{unequal} \approx e^{-\sigma^2} = e^{-(t/N)^2\sum_{i<j}\delta_{ij}^2}\end{equation}

\begin{figure}
    \centering
    \includegraphics[width=8cm,height=8cm,keepaspectratio]{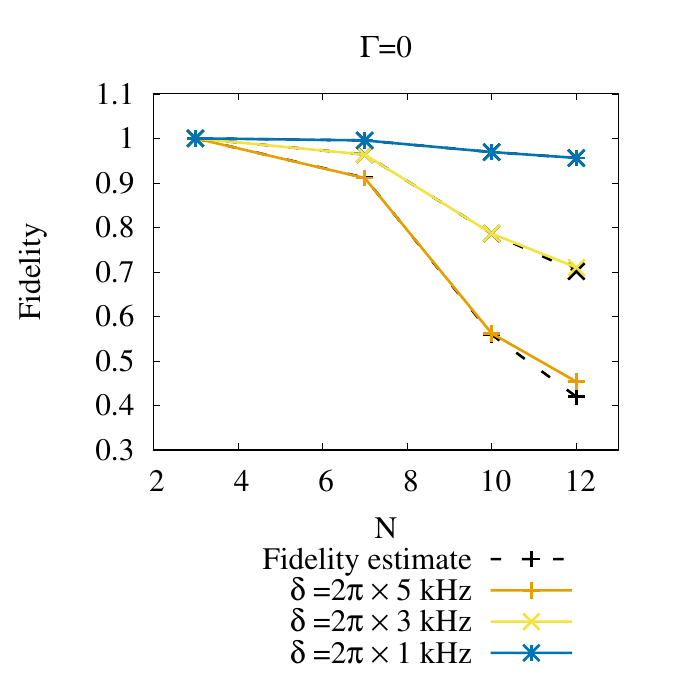}
    \caption{GHZ fidelity in the noiseless case, for varying detunings. Colored points are exact results (with infidelity due to unequal coupling), black points are estimates from (\ref{F approx}), and lines are included to guide the eye.}
    \label{fig:enter-label}
\end{figure}

Another way to express this is in terms of the variance of $J_{ij}$ over the ion-pairs $(i,j)$,
\begin{equation} \sigma^2(J_{ij}) = {N \choose 2}^{-1} \sum_{i < j} (J_{ij}-J)^2 = {N \choose 2}^{-1} \sum_{i < j}\delta_{i,j}^2 \end{equation}
and putting this into the previous relation gives  
\begin{equation} \label{F approx J variance}F_\text{unequal} \approx \exp\left(-{N \choose 2} \frac{t^2}{N^2} \sigma^2(J_{ij})\right)\end{equation}
For large $N$ we can simplify the expression, since ${ N \choose 2} \approx N^2/2$ and hence
\begin{equation} F_\text{unequal} \approx \exp\left[-t^2 \sigma^2(J_{ij})/2\right]\end{equation}

\subsection{Second derivation}

For the second derivation we use Euler's formula to expand $\exp(itN^{-1}\delta_{ij} \sigma^z_i\sigma^z_j) = \cos(tN^{-1}\delta_{ij}) + i\sin(tN^{-1}\delta_{ij})\sigma^z_i\sigma^z_j$ giving
\begin{equation} \label{F cos sin} F_\text{unequal} = |\bra{\psi_0} \left( \prod_{i<j} \cos(tN^{-1}\delta_{ij}) + i\sin(tN^{-1}\delta_{ij})\sigma^z_i\sigma^z_j \right) \ket{\psi_0}|^2\end{equation}
First note that $\bra{\psi_0}\sigma^z_i\ket{\psi_0}=0$ since $\ket{\psi_0}$ is an eigenstate of $\sigma^x_i$.  Hence, only terms in (\ref{F cos sin}) which do not contain any $\sigma^z_i$ operators will contribute to $F$.  We shall assume that the variations $\delta_{ij}$ are small so $c_{ij}= \cos(tN^{-1}\delta_{ij})\approx 1$ and $s_{ij}=\sin(tN^{-1}\delta_{ij}) \approx 0$, then perform an expansion in terms of the $c_{ij}$ and $s_{ij}$. The lowest-order term in this expansion is
\begin{equation} F_2 = \prod_{i<j} c_{ij}^2 = 1-\frac{t^2}{N^2}\sum_{i<j}\delta_{ij}^2 + \mathcal{O}(\delta^4) = e^{-(t/N)^2\sum_{i<j}\delta_{ij}^2}+ \mathcal{O}(\delta^4) \end{equation}
which agrees with (\ref{F approx}) up to corrections depending on products of four of the $\delta_{ij}$. The next lowest-order term includes all products of four sine terms in which each index appears twice, 
\begin{equation} F_4 = \sum_{k<l<m<n} 2(s_{kl}s_{lm}s_{mn}s_{nk} + s_{km}s_{kn}s_{ml}s_{nl} + s_{kl}s_{km}s_{ln}s_{mn}) \prod_{i<j | i,j \neq k,l,m,n} c_{ij}^2 \end{equation}
We approximate $F_\text{unequal} = F_2 + F_4$ for results in the main paper. 

\section{Spin squeezing and a benchmarking implementation of the  quantum approximate optimization algorithm}

Figure \ref{fig:squeezing Jt}(a) shows the optimized $Jt/N$ for spin squeezing from Fig.~4 of the main text, including the power-law fit associated with small effective magnetic fields in leakage.  

We find similarly small $Jt/N$ for a single-layer implementation of the quantum approximate optimization algorithm \cite{farhi2014quantum} in Fig.~\ref{fig:squeezing Jt}(b). For these calculations we begin with the standard expression for a single-layer QAOA 
\begin{equation} \ket{\psi} =e^{-i \beta B} e^{-i \gamma C}\ket{\psi_0} \end{equation}
with $\ket{\psi_0}=\ket{+}^n$, $B = \sum_i \sigma^x_i$, and an Ising Hamiltonian as the cost function 
\begin{equation} C = \frac{N}{J} \Hc = \frac{1}{J}\sum_{i<j} J_{ij} \sigma_i^z \sigma_j^z\end{equation}
where $J$ is the average of the $J_{ij}$, which is used to normalize the summands in $C$. The goal of QAOA is to prepare and measure $\ket{\psi}$ to obtain an approximate ground state of $C$.  

Typically $C$ encodes a combinatorial optimization problem such as an instance of Maximum-Cut. In previous work \cite{rajakumar2022generating}, it has been shown how all-to-all Ising evolution, generated from a native light-shift or M{\o}lmer-S{\o}rensen interaction in trapped ions, can be combined with targeted $\pi$-pulses to compile Ising interactions with arbitrary $J_{ij}$, which can represent a variety of interesting combinatorial problems $C$ for QAOA.  While this remains as an exciting frontier in the field, we are not able to model the sequences of $\pi$-pulses and global Ising evolutions, or the associated compilation sequences that are needed to solve computationally relevant problems, using our theory. Here we simply assess the ability to optimize with respect to the native Hamiltonian $\Hc_\text{se}$, similar to the previous benchmarking of Pagano {\it et al.} \cite{pagano2020quantum}.  The expected cost is 
\begin{align} \bra{\psi} C \ket{\psi} = \frac{1}{J}\sum_{ij} J_{ij} \bra{\psi_0} e^{i \gamma C} e^{i \beta B} \sigma_i^z \sigma_j^z e^{-i \beta B} e^{-i \gamma C} \ket{\psi_0} \nonumber\\
= \frac{1}{J}\sum_{i < j} J_{ij} \left(\cos^2(2\beta) \langle\sigma_i^z \sigma_j^z\rangle +\frac{\sin(4\beta)}{2} (\langle\sigma_i^y\sigma_j^z\rangle + \langle\sigma_i^z\sigma_j^y\rangle) + \sin^2(2\beta) \langle\sigma_i^y\sigma_j^y\rangle \right)\label{Expec C} \end{align}
where the expectation values in the final line are taken with respect to the Ising evolution $\exp(-i \gamma C)$ only. We evaluate $\langle C \rangle$ in noisy and noiseless cases in Fig.~\ref{fig:QAOA}, using $\gamma$ and $\beta$ optimized through a grid search over 101 evenly-spaced intervals with $0 \leq \gamma \leq \pi/\sqrt{N}$ and $-\pi/4 \leq \beta \leq \pi/4$ to determine the optimal $\langle C \rangle$.  The noisy $\langle C \rangle$ is hardly distinguishable from the noiseless case at this scale, though it has a slightly smaller numerical value. We attribute the small impact of scattering noise on $\langle C \rangle$ to the small $Jt/N$ generating negligible effective magnetic fields, along with similar dynamics and $\langle C \rangle$ generated from trajectories with and without leakage.

\begin{figure}
    \centering
    \includegraphics[height=7cm,width=15cm,keepaspectratio]{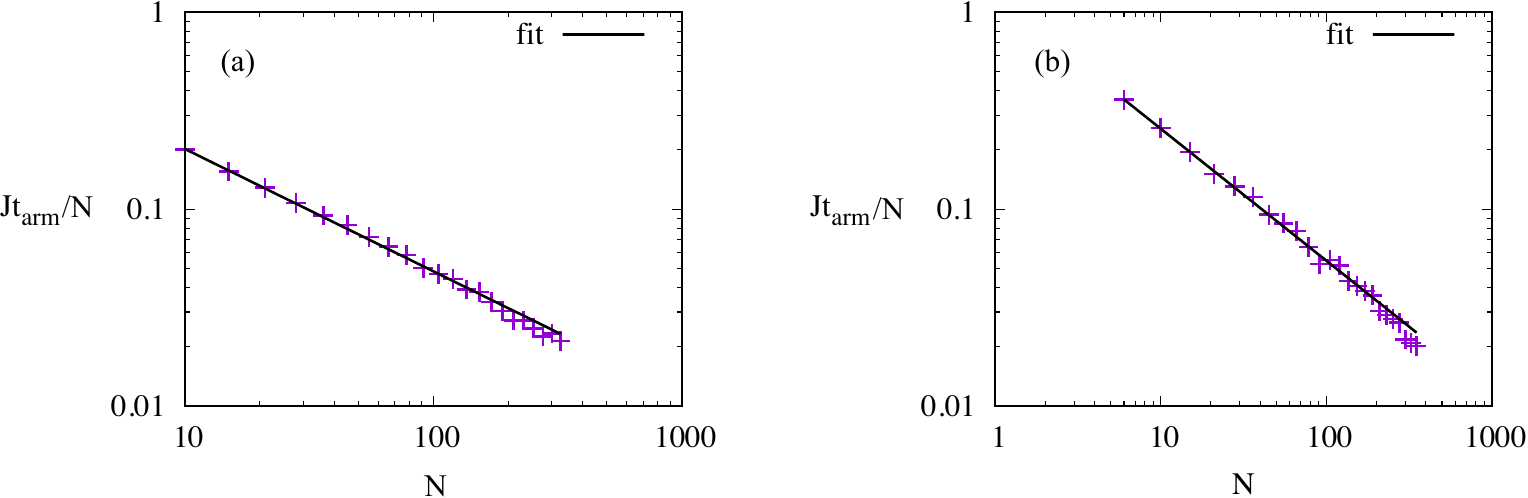}
    \caption{$Jt_\text{arm}/N$ for optimal (a) spin squeezing and (b) single-layer QAOA are well fit by $Jt_\text{arm}/N = aN^b$. For spin squeezing $a = 0.85 \pm 0.02$ and $b=-0.621 \pm 0.006$ while for QAOA $a=1.12 \pm 0.02$ and $b=-0.670 \pm 0.006$, with $\pm$ denoting the asymptotic standard error of the nonlinear least-squares fit.}
    \label{fig:squeezing Jt}
\end{figure}

\begin{figure}
    \centering
    \includegraphics[height=7cm,width=7cm,keepaspectratio]{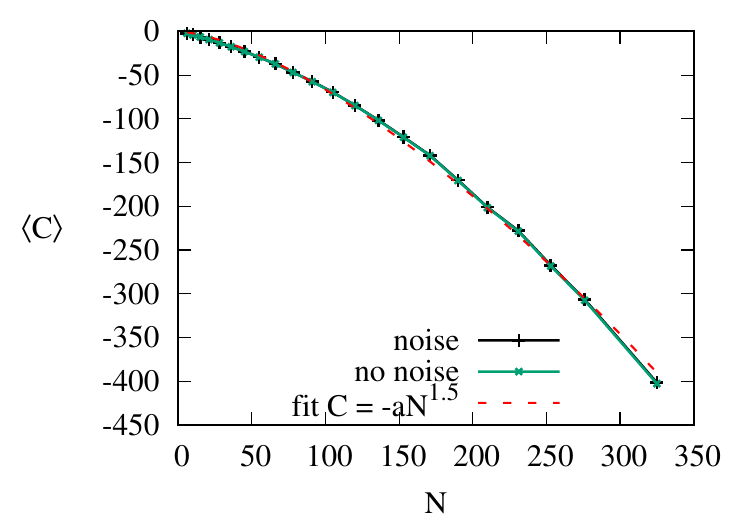}
    \caption{Optimal cost in QAOA in noisy and noiseless cases. The noisy results are fit by $\langle C \rangle = aN^{-1.5}$ with $a=0.0666 \pm 0.0004$.}
    \label{fig:QAOA}
\end{figure}

\end{widetext}

\bibliographystyle{unsrt}
\bibliography{refs}

\end{document}